\documentclass[lettersize,journal]{IEEEtran}
\usepackage{bbm}
\usepackage{subfigure}
\usepackage{amsmath,amsfonts}
\usepackage{algorithmic}
\usepackage{algorithm}
\usepackage{array}
\usepackage{textcomp}
\usepackage{stfloats}
\usepackage{url}
\usepackage{verbatim}
\usepackage{graphicx}
\usepackage{color}
\usepackage{cite}
\usepackage{amsthm}
\usepackage{amssymb}
\usepackage{cleveref}
\usepackage{makecell}
\usepackage{tikz}
\newcommand{\abs}[1]{\left\lvert #1\right\rvert}

\newcommand{\alphabet}{\mathbb A}

\newcommand{\eff}{\mathrm{eff}}
\usepackage{bbm}

\newcommand{\indic}[1]{\mathbbm{1}_{\left\{#1\right\}}}

\newcommand{\p}{\mathrm{p}}

\newcommand{\boldx}{\mathbf{x}}
\newcommand{\boldX}{\mathbf{X}}
\newcommand{\boldy}{\mathbf{y}}

\newcommand{\boldh}{\mathbf{h}}
\newcommand{\boldH}{\mathbf{H}}
\newcommand{\boldn}{\mathbf{n}}
\newcommand{\boldPhi}{\boldsymbol{\Phi}}
\newcommand{\boldtau}{\boldsymbol{\tau}}
\newcommand{\boldnu}{\boldsymbol{\nu}}

\newcommand{\rx}{\mathrm{rx}}

\newcommand{\taup}{\tau_{\text{p}}}
\newcommand{\nup}{\nu_{\text{p}}}
\newcommand{\dd}{\text{dd}}
\newcommand{\sinc}{\mathrm{sinc}}
\newcommand{\boldG}{\mathbf{G}}
\newcommand{\boldR}{\mathbf{R}}
\newcommand{\wrx}{w_{\text{rx}}}
\theoremstyle{plain}

\theoremstyle{definition}

\theoremstyle{remark}
\newtheorem{rem}{Remark}

\graphicspath{{Figures/}{../Figures/}}

\begin{document}
\title{Multiuser Zak-OTFS on the Uplink with Superimposed Spread-Pilots}
\author{Sai Pradeep Muppaneni and Ananthanarayanan Chockalingam \\ 
Department of ECE, Indian Institute of Science, Bangalore-560012
}

\maketitle
\begin{abstract}
In this paper, we consider the uplink of a multiuser Zak-OTFS system comprising users with heterogeneous delay-Doppler (DD) periods/frame sizes. Multiple access is achieved through time-frequency (TF) shifts that place the users in non-overlapping regions of the TF plane. Closed-form expressions for the effective DD domain channel between each user and the base station are derived for sinc and Gaussian pulse shaping filters. The inter-user interference (IUI) is shown to be negligible under the TF-shift-based multiple access, thereby decoupling the multiuser input-output relation (IOR) estimation problem into independent single-user estimation problems. For IOR estimation, a superimposed spread-pilot framework is employed. The spread-pilot sequence is obtained by applying FFT to a reshaped Zadoff-Chu sequence. To mitigate the pilot-data interference introduced by the superimposed spread-pilot, a DD dictionary-based IOR estimation scheme that iterates between IOR estimation and data detection is employed. Simulation results for a multiuser Zak-OTFS system demonstrate that the resulting IOR estimates achieve normalized mean-square error (NMSE) and bit error rate (BER) performances that closely match those of the corresponding single-user system. Furthermore, for sinc pulse shaping, the superimposed spread-pilot frame achieves higher spectral-efficiency compared to embedded pilot frame across a wide range of inter-user power ratios. For Gaussian pulse shaping, however, the embedded pilot frame achieves a higher spectral efficiency due to the combined effects of residual IUI and significant pilot-data interference in the case of superimposed spread-pilot. The robustness of the estimation framework to variations in channel power-delay profile and maximum Doppler shift is also demonstrated.
\end{abstract}

\section{Introduction}
\label{sec:intro}
Next-generation wireless communication networks are expected to support diverse and highly mobile scenarios, including non-terrestrial networks, high-speed train and vehicular communications, and aircraft-to-ground links. Such environments are characterized by channels with significant delay and Doppler spreads \cite{next_gen_ITU}. Orthogonal frequency division multiple access (OFDMA), which underpins current 4G and 5G systems, performs well in low/medium-mobility environments where the input-output relation (IOR) can be efficiently acquired. However, in high-mobility channels, the IOR in OFDMA becomes difficult to estimate and track. Acquiring the channel response across all subcarriers simultaneously is challenging in the presence of significant Doppler spreads (which is typical in high-mobility channels), resulting in increased pilot overhead and degraded performance \cite{otfs_WCNC}. Furthermore, OFDMA requires all users to operate with a common subcarrier spacing. Consequently, the presence of a single high-mobility user necessitates a larger subcarrier spacing for all users, leading to increased cyclic-prefix (CP) overhead and reduced spectral efficiency \cite{SE}.

Zak transform based OTFS modulation \cite{zakotfs_bits,zakotfs_pred}, referred to as Zak-OTFS, offers a promising alternative. Unlike multicarrier OTFS (MC-OTFS) \cite{mc_otfs_1}-\cite{mu_otfs_8}, which builds upon the existing OFDM-based framework, Zak-OTFS uses inverse Zak transform \cite{Zak_existance, zak_transform} to directly convert information symbols embedded in the DD domain into a continuous time domain (TD) signal. In Zak-OTFS, information symbols are carried by quasi-periodic pulses in the DD domain, whose TD realizations are pulse trains modulated by tones, referred to as pulsones. The IOR of Zak-OTFS is predictable and non-fading when the delay period exceeds the delay spread of the effective channel
and the Doppler period exceeds the Doppler spread of the effective channel, a condition referred to as the crystallization condition \cite{zakotfs_bits, zakotfs_pred}. In the crystalline regime, the taps of the effective DD-domain channel filter can be directly read off from the response to a single pilot pulsone, enabling simple and model-free channel acquisition with low overhead.

Closed-form expressions for the Zak-OTFS IOR in a single-user setting for sinc and Gaussian pulse shaping filters have been derived in \cite{closed_form}, revealing the role of pulse shaping filter in governing inter-symbol interference in the DD domain. It is shown in \cite{closed_form} that the sinc pulse yields superior BER performance compared to the Gaussian pulse, since the sinc pulse has zeros at the
sampling instants of adjacent DD bins, whereas the Gaussian pulse has nonzero power at those instants, resulting in a high inter-symbol interference. Further, the effect of a Gaussian-sinc (GS) composite pulse shape on the Zak-OTFS IOR and system performance has been studied in \cite{gsfilter}. For IOR estimation, an embedded pilot frame, in which a known pilot symbol surrounded by a guard region of zeros is placed at the center of the DD frame, has been used in
\cite{gsfilter}. While the embedded frame isolates the pilot from data symbols, it incurs a spectral
efficiency loss owing to the DD bins reserved exclusively for the pilot and its guard region. To address this, superimposed spread-pilot sequences that are spread over all DD bins of the frame and superimposed on the information symbols have been studied in \cite{isac, zadoff_spread}. While \cite{isac} employs a chirp-based spreading filter applied to a pilot pulsone to generate the superimposed spread pilot, \cite{zadoff_spread}
uses an FFT-based Zadoff--Chu (ZC) sequence as the spread pilot. It has been shown in \cite{isac, zadoff_spread} that the superimposed spread pilot not only utilizes all DD bins of the frame for data transmission but also reduces the peak-to-average power ratio (PAPR) of the resulting TD signal compared to the embedded frame. However, all of the above works are in the single-user setting, and the extension of Zak-OTFS to the multiuser setting is of significant practical importance.

A key practical strength of OFDMA is its ability to
allocate non-overlapping time-frequency (TF) resources to different users with flexible numerology, thereby enabling orthogonal multiple access with low inter-user interference \cite{goldsmith}. An analogous capability for Zak-OTFS in the uplink has been developed in
\cite{mu_paper}, where it is shown that by incorporating a TF shift into the transmit pulse
shaping filter of each user, the transmitted signals of all users can be confined to non-overlapping regions of the TF plane. The BS then applies individual matched filters to recover each user's signal. Since different users occupy distinct regions of the TF plane, the inter-user interference (IUI) is very small \cite{mu_paper}. 

In this paper, building on the multiuser Zak-OTFS framework of \cite{mu_paper} which uses embedded pilot frames, we investigate uplink multiuser Zak-OTFS 
with heterogeneous users using superimposed spread-pilot frames, which has not been reported. Superimposed spread-pilot frames offer the advantages of no pilot overhead and lower PAPR compared to embedded pilot frames. We consider sinc and Gaussian pulse shaping filters as they represent the canonical extremes of the orthogonalization (sinc) and localization (Gaussian) properties of pulse shapes which influence the IOR estimation and data detection performance. The key contributions of this work are summarized as follows.
\begin{itemize}
\item {\em Closed-form effective channel expressions:}
We derive closed-form expressions for the effective DD domain channel between each user and the BS for sinc and Gaussian pulse shaping filters. These expressions generalize the single-user expressions in \cite{closed_form} to multiuser setting with
heterogeneous DD periods/frame sizes and TF shifts for different users.

\item {\em Negligibility of IUI and user decoupling:}
Using the closed-form IOR expressions in an illustrative two-user experiment, we demonstrate that the IUI at each user's receiver is negligible under the TF-shift-based multiple access scheme. For example, in the absence of noise, the signal-to-interference power ratio (SIR) at User~1's receiver exceeds $15$~dB for Gaussian filter and $28$~dB for
sinc filter even when User~2 transmits at ten times
the power of User~1. This negligibility of IUI decouples the multiuser IOR estimation problem into 
independent single-user subproblems of identical complexity, which can be solved in parallel at the BS.

\item {\em Superimposed spread-pilot IOR estimation:}
We employ a superimposed spread-pilot framework for IOR estimation, where, for each user, a Zadoff-Chu (ZC) sequence  transformed by FFT along the Doppler axis \cite{zadoff_spread} is used as the spread-pilot sequence, which is superimposed on the data symbols of that user. To mitigate the pilot-data interference inherent in the superimposed pilot structure, we employ a DD dictionary-based iterative algorithm that alternates between IOR estimation and data detection. The proposed algorithm is shown to achieve near single-user performance with perfect CSI.

\item {\em Spectral efficiency~-~Embedded vs superimposed pilots:}
Comparing the spectral efficiency performance of embedded and superimposed spread-pilot frames, we show that, for sinc filter, the superimposed spread-pilot frame achieves higher spectral efficiency due to the elimination of pilot and guard bins. In contrast, for Gaussian filter, the embedded frame performs better due to the combined effects of residual IUI and pilot-data interference in the case of superimposed spread-pilot. These results highlight the preferred pilot structure for a given filter and the trade-off between pilot overhead and spectral efficiency.
\end{itemize}

The rest of this paper is organized as follows. In
Sec.~\ref{sec:sys_model}, we present the multiuser
Zak-OTFS system model. Closed-form expressions for IOR and noise covariance are derived in Sec. \ref{sec:closed_form}. Detailed derivations are moved to the Appendices. In Sec.~\ref{sec4}, we describe the superimposed spread-pilot frame construction at the transmitter and the IOR estimation algorithm at the receiver. Simulation results and discussions are presented in Sec.~\ref{sec:results}. Conclusions and future work are presented in Sec.~\ref{sec:conclusion}. 

\subsubsection*{Notations}
Upper case and lowercase boldface letters denote matrices and vectors, respectively. $(\cdot)^H$
and $(\cdot)^{-1}$ denote conjugate transpose and matrix inverse, respectively. $\mathrm{vec}(\cdot)$ denotes the vectorization operator that stacks the columns of a matrix into a column vector. $\|\cdot\|_F$ denotes the Frobenius norm of a matrix. $\mathbb{C}^{m\times n}$ denotes the set of complex matrices of size $m\times n$. $\mathbf{0}_{m\times n}$ denotes the $m\times n$ all-zeros matrix. $\mathbb{E}[\cdot]$ denotes the expectation operator. $\lfloor\cdot\rfloor$ and $\lceil\cdot\rceil$ denote the floor and ceiling operators, respectively, and $a\bmod b$ denotes the remainder of $a$ divided by $b$. $\delta(\cdot)$ denotes the Dirac delta function and $\indic{\cdot}$ denotes the indicator function. $\sinc(x)\triangleq\frac{\sin(\pi x)}{\pi x}$ denotes the normalized sinc function. $\inf(\cdot)$ and $\sup(\cdot)$ denote the infimum and supremum of a set, respectively. For a matrix $\mathbf{A}$,
$\mathbf{A}[i,j]$ denotes its $(i,j)$th entry. For a vector $\mathbf{a}$, $\mathbf{a}[i]$ denotes its $i$th entry. $|\mathcal{S}|$ denotes the cardinality of a set $\mathcal{S}$. $\mathbb{R}^+\backslash\{0\}$ denotes the set of positive real numbers excluding zero.

\section{Uplink multiuser Zak-OTFS system model}
\label{sec:sys_model}
\begin{figure*}[t]
    \centering  \includegraphics[width=17cm,height=7.5cm]
    {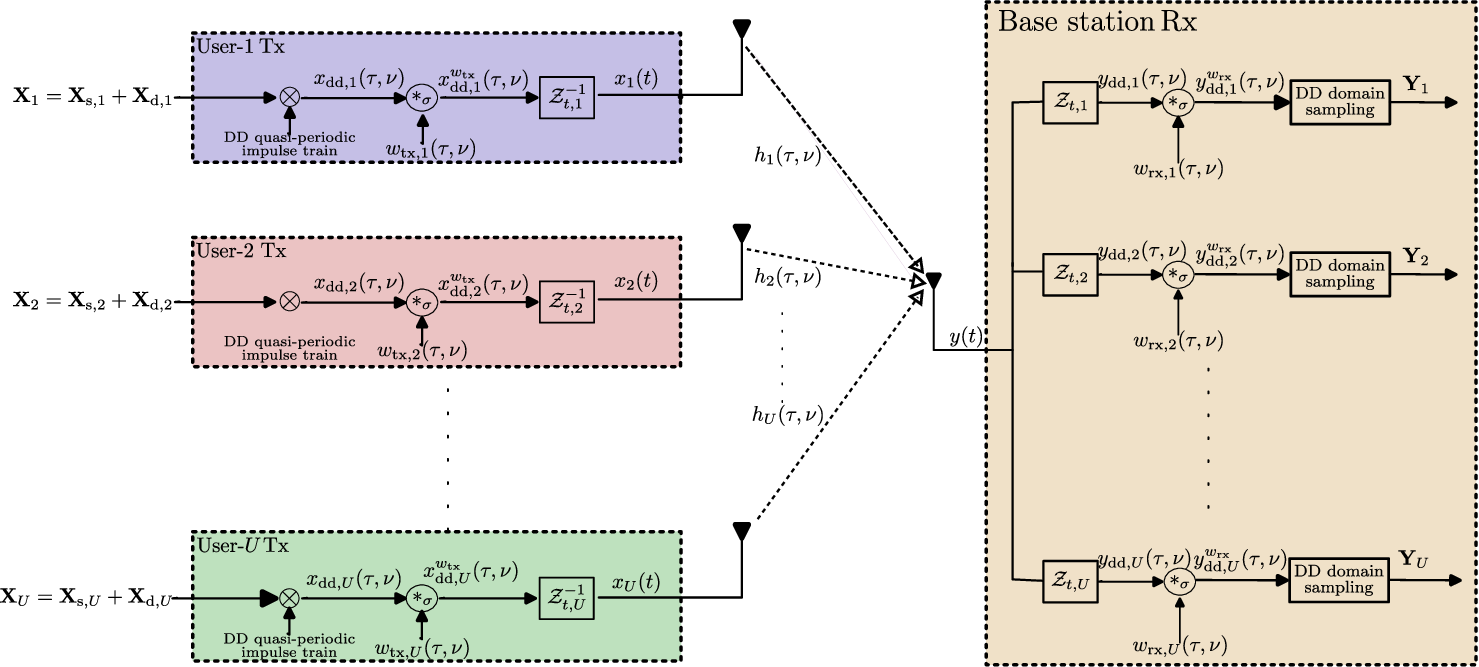}
    \caption{Block diagram of the uplink multiuser Zak-OTFS system.}
    \label{fig:block_diag}
\end{figure*}
Consider the uplink of a multiuser Zak-OTFS system with $U$ users and a single base station (BS), as illustrated in Fig.~\ref{fig:block_diag}. The $u$th user, $u \in \{1, 2, \ldots, U\}$, operates with $M_u$ delay bins and $N_u$ Doppler bins. The 
bandwidth and frame duration of the $u$th user are given by $B_u = M_u \nu_{\text{p},u}$ and $T_u = N_u \tau_{\text{p},u}$, respectively, where $\tau_{\text{p},u}$ and $\nu_{\text{p},u}$ denote the delay period and Doppler period of the $u$th user, respectively, satisfying $\tau_{\text{p},u} \nu_{\text{p},u} = 1.$ The $u$th user transmit signal matrix is $\mathbf{X}_u \in \mathbb{C}^{M_u \times N_u}$, whose elements are placed on the DD grid at locations $\left(\frac{k \tau_{\text{p},u}}{M_u},\, \frac{l \nu_{\text{p},u}}{N_u}\right)$, 
where $k \in \{0, 1, \ldots, M_u - 1\}$, $l \in \{0, 1, \ldots, N_u - 1\}$. The DD domain signal $x_{\mathrm{dd},u}(\tau, \nu)$ corresponding to 
$\mathbf{X}_u$ is constructed as
\begin{align}
    x_{\mathrm{dd},u}(\tau, \nu)
    &= \sum_{r,s\in\mathbb{Z}}
    \mathbf{X}_u[r \bmod M_u,\, s \bmod N_u]\,
    e^{j2\pi \lfloor \frac{r}{M_u} \rfloor \frac{s}{N_u}}
    \nonumber \\
    &\quad 
    \delta\!\left(\tau - r\frac{\tau_{\text{p},u}}{M_u}\right)
    \delta\!\left(\nu - s\frac{\nu_{\text{p},u}}{N_u}\right),
    \label{eq:dd_signal}
\end{align}
which is quasi-periodic in $(\tau, \nu)$. The DD signal $x_{\mathrm{dd},u}(\tau, \nu)$ is passed through a transmit filter $w_{\mathrm{tx},u}(\tau, \nu)$ via twisted convolution, denoted $\ast_\sigma$, to limit the bandwidth and time duration of the transmitted signal. The filtered DD signal $x_{\mathrm{dd},u}^{w_{\mathrm{tx}}}(\tau, \nu)$ is given by
\begin{align}
    x_{\mathrm{dd},u}^{w_{\mathrm{tx}}}(\tau, \nu)
    &= w_{\mathrm{tx},u}(\tau, \nu) \ast_\sigma x_{\mathrm{dd},u}(\tau, \nu)
    \nonumber \\
    &= \iint w_{\mathrm{tx},u}(\tau', \nu')\,
    x_{\mathrm{dd},u}(\tau - \tau',\, \nu - \nu')
    \nonumber \\
    &\quad 
    e^{j2\pi \nu'(\tau - \tau')}\, d\tau'\, d\nu'.
    \label{eq:twisted_conv}
\end{align}
The transmit filter for the $u$th user is chosen as
\begin{equation}
    w_{\mathrm{tx},u}(\tau, \nu)
    = w_{B_u}(\tau)\, w_{T_u}(\nu)\,
    e^{j2\pi(\nu_u \tau - \tau_u \nu)},
    \label{eq:tx_filter}
\end{equation}
where $w_B(\tau)$ and $w_T(\nu)$ are prototype pulse shapes along the delay and Doppler dimensions, respectively, and the phase term $e^{j2\pi(\nu_u \tau - \tau_u \nu)}$ introduces a shift of $(\tau_u, \nu_u)$ in the time-frequency (TF) plane \cite{mu_paper}.

\begin{figure}
    \centering
    \subfigure[Without TF shift.]
        {\label{fig:demo_1}
         \includegraphics[height=0.35\linewidth]{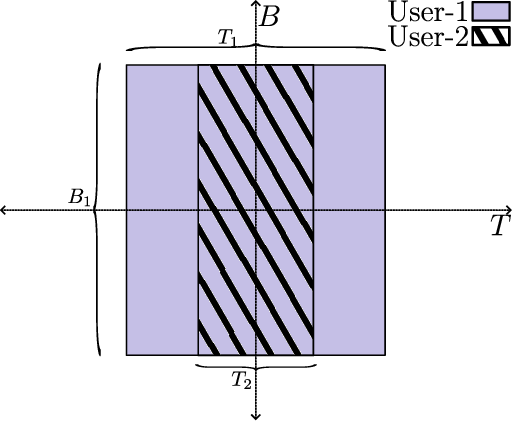}}
    \subfigure[With TF shift.]
        {\label{fig:demo_2}
         \includegraphics[height=0.35\linewidth]{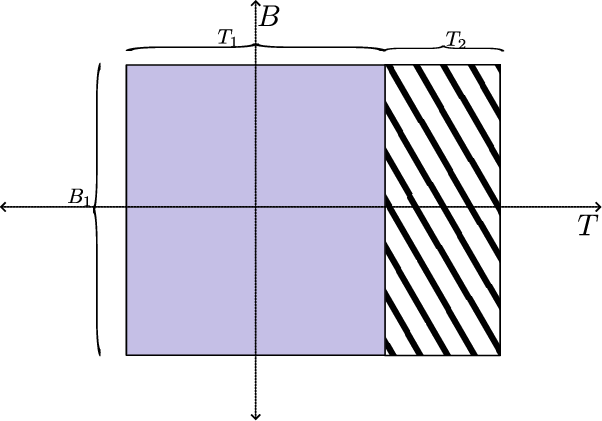}}
    \caption{Illustration of TF support regions for a two-user system with and without TF shift.}
    \label{fig:tf_shifts}
\end{figure}

\begin{rem}
The phase term $ e^{j2\pi(\nu_u \tau - \tau_u \nu)}$ in \eqref{eq:tx_filter} serves to place 
different users in non-overlapping regions of the TF plane, thereby enabling simultaneous uplink transmission with negligible IUI. This is illustrated in Fig.~\ref{fig:tf_shifts} for a two-user example. Without TF shifts, the support regions of the two users overlap (Fig.~\ref{fig:demo_1}). By introducing a shift of $(0, \frac{T_1 + T_2}{2})$ via the phase term $e^{-j2\pi \frac{T_1+T_2}{2} \nu}$ in the second user's filter, the two support regions are rendered non-overlapping (Fig.~\ref{fig:demo_2}), facilitating simultaneous communication.
\end{rem}

The filtered DD signal $x_{\mathrm{dd},u}^{w_{\mathrm{tx}}}(\tau, \nu)$ is converted to TD via the inverse Zak transform (IZT) as
\begin{align}
    x_u(t)
    &= \mathcal{Z}^{-1}_{t,u}
       \!\left(x_{\mathrm{dd},u}^{w_{\mathrm{tx}}}(\tau, \nu)\right)
    \nonumber \\
    &= \sqrt{\tau_{\text{p},u}}
       \int_0^{\nu_{\text{p},u}}
       x_{\mathrm{dd},u}^{w_{\mathrm{tx}}}(t, \nu)\, d\nu,
    \label{eq:izt}
\end{align}
where $\mathcal{Z}^{-1}_{t,u}(\cdot)$ denotes the IZT operator for the $u$th user. The doubly-dispersive channel of the $u$th user is modeled as a superposition of $P_u$ propagation paths, given by
\begin{equation}
    h_u(\tau, \nu)
    = \sum_{i=1}^{P_u}
      h_{i,u}\,
      \delta(\tau - \tau_{i,u})\,
      \delta(\nu - \nu_{i,u}),
    \label{eq:channel}
\end{equation}
where $h_{i,u}$, $\tau_{i,u}$, and $\nu_{i,u}$ denote the complex path gain, delay, and Doppler shift of the $i$th path of the $u$th user, respectively. The signal received at the BS is a superposition of the time-delayed and Doppler-shifted transmitted signals from all $U$ users, corrupted by AWGN $n(t)$, which is given by
\begin{align}
    y(t)
    &= \sum_{u=1}^{U} \sum_{i=1}^{P_u}
       h_{i,u}\, x_u(t - \tau_{i,u})\,
       e^{j2\pi \nu_{i,u}(t - \tau_{i,u})}
       + n(t).
    \label{eq:rx_signal}
\end{align}
The BS applies Zak transform on $y(t)$ to obtain the $u$th user's signal as  
\begin{equation}
\begin{aligned}
     y_{\dd,u}(\tau,\nu)&=\mathcal{Z}_{t,u}\left(y(t)\right) \\ &=\sqrt{\tau_{\p,u}}\sum_{q\in\mathbb{Z}}y(\tau+q\tau_{\p,u})e^{-j2\pi q\frac{\nu}{\nu_{\p,u}}},
\end{aligned}
\end{equation}
and filters the resulting DD domain signal using the $u$th user's receive filter $w_{\mathrm{rx},u}(\tau,\nu)
    \triangleq
    w_{\mathrm{tx},u}^{*}(-\tau,-\nu)\,
    e^{j2\pi\tau\nu},
    \quad u \in \{1,\ldots,U\}.$
The filtered DD domain signal of the $u$th user is then given by
\begin{eqnarray}
y_{\mathrm{dd},u}^{w_{\mathrm{rx}}}(\tau,\nu)
& \hspace{-2mm} = & \hspace{-2mm}
w_{\mathrm{rx},u}(\tau,\nu)
\ast_{\sigma}
\left(y_{\dd,u}(\tau,\nu)\right)
\nonumber \\
& \hspace{-2mm} = & \hspace{-2mm}
w_{\mathrm{rx},u}(\tau,\nu)
\ast_{\sigma}
\Bigg(
\sum_{v=1}^{U}
h_v(\tau,\nu)
\ast_{\sigma}
x_{\mathrm{dd},v}^{w_{\mathrm{tx}}}(\tau,\nu)
\notag \\
& \hspace{-2mm} & \hspace{-2mm} + \
n_{\mathrm{dd},u}(\tau,\nu)\Bigg)
\nonumber\\
& \hspace{-2mm} = & \hspace{-2mm}
\sum_{v=1}^{U}
\underbrace{
\Big(
w_{\mathrm{rx},u}(\tau,\nu)
\ast_{\sigma}
h_v(\tau,\nu)
\ast_{\sigma}
w_{\mathrm{tx},v}(\tau,\nu)
\Big)
}_{\triangleq\, h_{\mathrm{eff},u,v}(\tau,\nu)}
\notag \\
& \hspace{-2mm} & \hspace{-2mm} 
\ast_{\sigma}
x_{\mathrm{dd},v}(\tau,\nu)
+
n_{\mathrm{dd},u}^{w_{\mathrm{rx}}}(\tau,\nu),
\label{eq:ydd_filtered}
\end{eqnarray}
where $n_{\mathrm{dd},u}(\tau,\nu)$ is the additive noise in the DD domain, $n_{\mathrm{dd},u}^{w_{\mathrm{rx}}}(\tau,\nu) \triangleq w_{\mathrm{rx},u}(\tau,\nu)
\ast_{\sigma} n_{\mathrm{dd},u}(\tau,\nu)$
denotes the filtered noise in the DD domain, and
$h_{\mathrm{eff},u,v}(\tau,\nu)$ denotes the effective DD domain channel between the $v$th user's transmitter and the $u$th user's receive filter at the BS. We note that \eqref{eq:ydd_filtered} admits a clean decomposition into a desired signal term and an IUI
term as follows:
\begin{eqnarray}
\hspace{-4mm}
    y_{\mathrm{dd},u}^{w_{\mathrm{rx}}}(\tau,\nu)
    & \hspace{-2mm} = & \hspace{-2mm} 
    \underbrace{
        h_{\mathrm{eff},u,u}(\tau,\nu)
        \ast_{\sigma}
        x_{\mathrm{dd},u}(\tau,\nu)
    }_{\text{desired signal}}
    \nonumber\\
    & \hspace{-14mm} & \hspace{-14mm} 
    + \underbrace{
        \sum_{\substack{v=1\\v\neq u}}^{U}
        h_{\mathrm{eff},u,v}(\tau,\nu)
        \ast_{\sigma}
        x_{\mathrm{dd},v}(\tau,\nu)
    }_{\text{IUI from other users}}
    + \
    n_{\mathrm{dd},u}^{w_{\mathrm{rx}}}(\tau,\nu).
    \label{eq:ydd_decomposed}
\end{eqnarray}
The continuous DD domain signal
$y_{\mathrm{dd},u}^{w_{\mathrm{rx}}}(\tau,\nu)$ is sampled at the DD grid points
$\left(\frac{k'\tau_{\text{p},u}}{M_u},\,\frac{l'\nu_{\text{p},u}}{N_u}\right)$,
$k' \in \{0,1,\ldots,M_u-1\}$,
$l' \in \{0,1,\ldots,N_u-1\}$,
to obtain the received symbol matrix
$\mathbf{Y}_u \in \mathbb{C}^{M_u \times N_u}$, with entries
\begin{align}
    \mathbf{Y}_u[k',l']
    &\triangleq
    y_{\mathrm{dd},u}^{w_{\mathrm{rx}}}(\tau,\nu)
    \bigg|_{\tau = \frac{k'\tau_{\text{p},u}}{M_u},\;
            \nu  = \frac{l'\nu_{\text{p},u}}{N_u}}.
    \label{eq:Yu_samples}
\end{align}
For notational convenience, define
\begin{align}
    \mathbf{Y}_{u,v}[k',l']
    &\triangleq
    \left.
    \Big(
        h_{\mathrm{eff},u,v}(\tau,\nu)
        \ast_{\sigma}
        x_{\mathrm{dd},v}(\tau,\nu)
    \Big)
    \right|_{\tau = \frac{k'\tau_{\text{p},u}}{M_u},\;
              \nu  = \frac{l'\nu_{\text{p},u}}{N_u}},
    \label{eq:Yuv_def}
\end{align}
so that
\begin{align}
    \mathbf{Y}_u[k',l']
    &=
    \sum_{v=1}^{U}
    \mathbf{Y}_{u,v}[k',l']+n_{\dd,u}^{\wrx}[k',l'],
    \label{eq:Yu_sum}
\end{align}
where $n_{\dd,u}^{\wrx}[k',l']=n_{\dd,u}^{w_{\rx}}(\tau,\nu)|_{\tau = \frac{k'\tau_{\text{p},u}}{M_u},\;
              \nu  = \frac{l'\nu_{\text{p},u}}{N_u}}$
Substituting the DD domain signal expression from
\eqref{eq:dd_signal} into \eqref{eq:Yuv_def}, and
using the index substitution $r = k + nM_v$, $s = l + mN_v$, $k \in \{0,\ldots,M_v-1\}$, $l \in \{0,\ldots,N_v-1\}$, $m,n \in \mathbb{Z}$, we obtain IOR as \eqref{eq:Yuv_expanded} given at the top of the next page \cite{mu_paper}.
\begin{figure*}[t]
\begin{align}
    \mathbf{Y}_{u,v}[k',l']
    &=
    \sum_{m,n \in \mathbb{Z}}
    \sum_{k=0}^{M_v-1}
    \sum_{l=0}^{N_v-1}
    h_{\mathrm{eff},u,v}
    \!\left(
        \frac{k'\tau_{p,u}}{M_u}
        - n\tau_{p,v}
        - \frac{k\tau_{p,v}}{M_v},\;
        \frac{l'\nu_{p,u}}{N_u}
        - m\nu_{p,v}
        - \frac{l\nu_{p,v}}{N_v}
    \right)
    \mathbf{X}_v[k,l]\nonumber\\
    &\quad \ \ 
    e^{j2\pi nl/N_v}
    e^{j2\pi
        \left(
            \frac{l'\nu_{p,u}}{N_u}
            - \frac{l\nu_{p,v}}{N_v}
            - m\nu_{p,v}
        \right)
        \left(
            \frac{k\tau_{p,v}}{M_v}
            + n\tau_{p,v}
        \right)
    }.
    \label{eq:Yuv_expanded}
\end{align}    
\hrule
\end{figure*}

We now cast the IOR into a compact matrix-vector form. Let $\mathbf{x}_v \triangleq \mathrm{vec}(\mathbf{X}_v) \in \mathbb{C}^{M_v N_v \times 1}$ and $\mathbf{y}_{u,v} \triangleq \mathrm{vec}(\mathbf{Y}_{u,v}) \in \mathbb{C}^{M_u N_u \times 1}$, $\forall~u,v\in\{1,\cdots,U\}$. It then follows from \eqref{eq:Yuv_expanded} that
\begin{equation}
    \mathbf{y}_{u,v}
    = \mathbf{H}_{u,v}\, \mathbf{x}_v,
    \label{eq:yuv_matrix}
\end{equation}
where $\mathbf{H}_{u,v} \in \mathbb{C}^{M_u N_u \times M_v N_v}$ is the IOR matrix whose $(l'M_u + k',\, lM_v + k)$th entry is given in \eqref{eqn:H_u,v} at the top of the next page.
\begin{figure*}[t]
\begin{align}
    \mathbf{H}_{u,v}[l'M_u+k',\, lM_v+k]
    &=
    \sum_{m,n \in \mathbb{Z}}
    h_{\mathrm{eff},u,v}
    \!\left(
        \frac{k'\tau_{p,u}}{M_u}
        - n\tau_{p,v}
        - \frac{k\tau_{p,v}}{M_v},\;
        \frac{l'\nu_{p,u}}{N_u}
        - m\nu_{p,v}
        - \frac{l\nu_{p,v}}{N_v}
    \right)
    e^{j2\pi nl/N_v}
    \nonumber\\
    &\quad \ \
    e^{j2\pi
        \left(
            \frac{l'\nu_{p,u}}{N_u}
            - \frac{l\nu_{p,v}}{N_v}
            - m\nu_{p,v}
        \right)
        \left(
            \frac{k\tau_{p,v}}{M_v}
            + n\tau_{p,v}
        \right)
    }.
    \label{eqn:H_u,v}
\end{align}
\hrule
\end{figure*}
Combining contributions from all $U$ users, the overall received signal vector $\mathbf{y}_u \triangleq \mathrm{vec}(\mathbf{Y}_u) \in \mathbb{C}^{M_u N_u \times 1}$ is given by
\begin{align}
    \mathbf{y}_u
    &=
    \sum_{v=1}^{U}
    \mathbf{H}_{u,v}\, \mathbf{x}_v
    + \mathbf{n}_u
    \nonumber\\
    &=
    \underbrace{
        \mathbf{H}_{u,u}\, \mathbf{x}_u
    }_{\text{desired signal}}
    +
    \underbrace{
        \sum_{\substack{v=1\\v\neq u}}^{U}
        \mathbf{H}_{u,v}\, \mathbf{x}_v
    }_{\text{IUI from other users}}
    + \ \mathbf{n}_u,
    \label{eq:yu_final}
\end{align}
where $\mathbf{n}_{u}\in\mathbb{C}^{M_uN_u\times1}:\  \mathbf{n}_{u}[l'M_u+k']=n_{\dd,u}^{\wrx}[k',l']$ denotes the noise vector at the $u$th user's receiver.

We consider sinc and Gaussian pulse shapes for $w_{B_u}(\tau)$ and $w_{T_u}(\nu)$. For the sinc pulse,
\[
w_{B_u}(\tau)=\sqrt{B_u}\sinc(B_u\tau), \quad w_{T_u}(\nu)=\sqrt{T_u}\sinc(T_u\nu).
\]
For the Gaussian pulse,
\[
w_{B_u}(\tau)=\left(\frac{2\alpha_\tau B_u^2}{\pi}\right)^{1/4} e^{-\alpha_\tau B_u^2\tau^2},\]and\[ w_{T_u}(\nu)=\left(\frac{2\alpha_\nu T_u^2}{\pi}\right)^{1/4} e^{-\alpha_\nu T_u^2\nu^2},
\]
where $\alpha_\tau=\alpha_\nu=1.584$ are chosen such that approximately $99\%$ of the pulse energy is contained within the signaling bandwidth and frame duration, respectively. 

\section{Closed-form expressions for IOR} 
\label{sec:closed_form}
In this section, we derive closed-form expressions for the effective channel coefficients $h_{\eff,u,v}(\tau,\nu)$ and noise covariance for the cases of 
sinc and Gaussian pulse shaping filters. 

\subsection{Closed-form expressions for $h_{\eff,u,v}(\tau,\nu)$}
{\em Sinc filter:} The closed-form expression for $h_{\eff,u,v}(\tau,\nu)$ for sinc filter is derived as
\begin{eqnarray}
h_{\eff,u,v}(\tau,\nu)
& \hspace{-2mm} = & \hspace{-2mm}
\sum_{i=1}^{P_v}
h_{i,v}
e^{j2\pi\nu_v(\tau-\tau_{i,v})}
e^{j2\pi\nu_{i,v}(\tau+\tau_v-\tau_{i,v})}
\nonumber \\
& \hspace{-17mm} & \hspace{-17mm}
e^{-j2\pi\nu\tau_v}
\tfrac{(b^{+}-b^{-})(t^{+}-t^{-})}{\sqrt{B_uB_vT_uT_v}}
\indic{\mathcal{B}_u\cap\mathcal{B}_v\neq\emptyset}
\indic{\mathcal{T}_u\cap\mathcal{T}_v\neq\emptyset}
\nonumber \\
& \hspace{-17mm} & \hspace{-17mm} 
\sinc((\tau_{i,v}-\tau)(b^{+}-b^{-}))
\sinc((\nu_{i,v}-\nu)(t^{+}-t^{-})) 
\nonumber \\
& \hspace{-17mm} & \hspace{-17mm}
e^{-j\pi(\tau_{i,v}-\tau)(b^{-}+b^{+})}
e^{-j\pi(\nu_{i,v}-\nu)(t^{-}+t^{+})},
\label{eq:heff_sinc}
\end{eqnarray}
where
\begin{align*}
\mathcal{B}_u &= \left[-\frac{B_u}{2}+(\nu_u-\nu_v-\nu_{i,v}), \, \frac{B_u}{2}+(\nu_u-\nu_v-\nu_{i,v})\right), \\ \mathcal{B}_v& = \left[-\frac{B_v}{2}, \, \frac{B_v}{2}\right),\quad \mathcal{T}_v = \left[-\frac{T_v}{2}, \, \frac{T_v}{2}\right), \\
\mathcal{T}_u &= \left[-\frac{T_u}{2}+(\tau-\tau_u+\tau_v), \, \frac{T_u}{2}+(\tau-\tau_u+\tau_v)\right),\\
b^{-}&=\inf(\mathcal{B}_u\cap\mathcal{B}_v),\quad b^{+}=\sup(\mathcal{B}_u\cap\mathcal{B}_v),\\
t^{-}&=\inf(\mathcal{T}_u\cap\mathcal{T}_v),\quad t^{+}=\sup(\mathcal{T}_u\cap\mathcal{T}_v).
\end{align*}
The derivation of (\ref{eq:heff_sinc}) is given in Appendix \ref{appx_A}.

{\em Gaussian filter:} The closed-form expression for $h_{\eff,u,v}(\tau,\nu)$ for Gaussian filter is derived as
\begin{eqnarray}
h_{\eff,u,v}(\tau,\nu)
& \hspace{-2mm} = & \hspace{-2mm}
\sum_{i=1}^{P_v}
h_{i,v}
\tfrac{2\sqrt{B_uB_vT_uT_v}}{\sqrt{(B_u^2+B_v^2)(T_u^2+T_v^2)}}
e^{j2\pi\nu_v(\tau-\tau_{i,v})}
\nonumber \\ 
& \hspace{-24mm} & \hspace{-24mm} 
e^{j2\pi\nu_{i,v}(\tau+\tau_v-\tau_{i,v})}
e^{-j2\pi\nu\tau_v}
e^{-\frac{\alpha_\tau B_u^2B_v^2}{B_u^2+B_v^2}(\tau-\tau_{i,v})^2}
\nonumber \\
& \hspace{-24mm} & \hspace{-24mm}
e^{-\frac{\pi^2(\nu_u-\nu_v-\nu_{i,v})^2}{\alpha_\tau(B_u^2+B_v^2)}}
e^{-\frac{\alpha_\nu T_u^2T_v^2}{T_u^2+T_v^2}(\nu-\nu_{i,v})^2}
e^{-\frac{\pi^2(\tau-\tau_u+\tau_v)^2}{\alpha_\nu(T_u^2+T_v^2)}}
\nonumber \\
& \hspace{-24mm} & \hspace{-24mm}
e^{j2\pi\frac{B_v^2}{B_u^2+B_v^2}(\tau-\tau_{i,v})(\nu_u-\nu_v-\nu_{i,v})}
e^{j2\pi\frac{T_v^2}{T_u^2+T_v^2}(\nu-\nu_{i,v})(\tau-\tau_u+\tau_v)}. \nonumber \\
\label{eq:heff_gaussian}
\end{eqnarray}
The derivation of (\ref{eq:heff_gaussian}) is given in Appendix \ref{appx_B}.

\subsection{Closed-form expressions for noise covariance}
{\em Sinc filter:} The closed-form expression for noise covariance for sinc filter is derived as
\begin{eqnarray}
\mathbb{E}\!\left[\, n_{\mathrm{dd},u}^{w_{rx}}[k_1,l_1]\; n_{\mathrm{dd},u}^{w_{rx}*}[k_2,l_2]\,\right]
& \hspace{-2mm} = & \hspace{-2mm} N_0 \frac{\tau_{\mathrm{p},u}}{T_u} \!\!\!
   \sum_{q_1=\left\lceil -\frac{N_u}{2}-\frac{k_1}{M_u}+\frac{\tau_u}{\tau_{\mathrm{p},u}}\right\rceil}
        ^{\left\lfloor \frac{N_u}{2}-\frac{k_1}{M_u}+\frac{\tau_u}{\tau_{\mathrm{p},u}}\right\rfloor}
        \nonumber \\
        & \hspace{-48mm} & \hspace{-48mm}
   \sum_{q_2=\left\lceil -\frac{N_u}{2}-\frac{k_2}{M_u}+\frac{\tau_u}{\tau_{\mathrm{p},u}}\right\rceil}
        ^{\left\lfloor \frac{N_u}{2}-\frac{k_2}{M_u}+\frac{\tau_u}{\tau_{\mathrm{p},u}}\right\rfloor}
        \hspace{-8mm}
   \sinc\!\left(B_u\Big((k_2-k_1)\tfrac{\tau_{\mathrm{p},u}}{M_u}+(q_2-q_1)\tau_{\mathrm{p},u}\Big)\right)
   \nonumber \\
   & \hspace{-48mm} & \hspace{-48mm}
   e^{\,j\frac{2\pi}{N_u}(q_2 l_2-q_1 l_1)}e^{j2\pi\nu_u\left(\frac{(k_1-k_2)\tau_{\mathrm{p},u}}{M_u}+(q_1-q_2)\tau_{\mathrm{p},u}\right)}.
   \label{eqn:sinc_noise_cov1}
\end{eqnarray}
The derivation of (\ref{eqn:sinc_noise_cov1}) is given in Appendix \ref{appx_C}.

{\em Gaussian filter:}
The closed-form expression for noise covariance for Gaussian filter is derived as
\begin{eqnarray}
\mathbb{E}\!\left[
n_{\dd,u}^{w_{\mathrm{rx}}}[k_1,l_1]\,
n_{\dd,u}^{w_{\mathrm{rx}}*}[k_2,l_2]
\right] & \hspace{-2mm} = & \hspace{-2mm}
N_0
\sqrt{\tfrac{2\pi}{\alpha_\nu T_u^2}}\,
\tau_{\mathrm{p},u}
\nonumber \\
& \hspace{-46mm} & \hspace{-46mm}
\sum_{q_1\in\mathbb{Z}}
\sum_{q_2\in\mathbb{Z}}
e^{-j\frac{2\pi}{N_u}(q_1l_1-q_2l_2)}
e^{-\frac{\pi^2}{\alpha_\nu T_u^2}
\left(
\frac{q_1}{\nu_{\mathrm{p},u}}
+\frac{k_1\tau_{\mathrm{p},u}}{M_u}
-\tau_u
\right)^2}
\nonumber \\
& \hspace{-46mm} & \hspace{-46mm}
e^{-\frac{\pi^2}{\alpha_\nu T_u^2}
\left(
\frac{q_2}{\nu_{\mathrm{p},u}}
+\frac{k_2\tau_{\mathrm{p},u}}{M_u}
-\tau_u
\right)^2}
e^{j2\pi\nu_u\left(\frac{(k_1-k_2)\tau_{\mathrm{p},u}}{M_u}
+
(q_1-q_2)\tau_{\mathrm{p},u}\right)}
\nonumber \\
& \hspace{-46mm} & \hspace{-46mm}
e^{-\frac{\alpha_\tau B_u^2}{2}
\left(\frac{(k_1-k_2)\tau_{\mathrm{p},u}}{M_u}
+
(q_1-q_2)\tau_{\mathrm{p},u}\right)^2}.
\label{eqn:gauss_noise_cov1}
\end{eqnarray}
The derivation of (\ref{eqn:gauss_noise_cov1}) is given in Appendix \ref{appx_D}.
\section{Transmit Frame and IOR Estimation}
\label{sec4}
In this section, we describe the construction of the superimposed spread-pilot frame at the transmitter and the IOR estimation algorithm at the receiver.

\subsection{Superimposed spread-pilot frame}
\label{sec:spread_pilot}
For the purpose of IOR estimation, the transmit 
frame of the $u$th user,  $\boldX_u$, is formed by 
a superposition of a spread-pilot and data symbols, where the spread-pilot component occupies all the $M_uN_u$ DD bins of the $u$th user's frame along with $M_uN_u$ data symbols. The pilot matrix
$\boldX_{\text{s},u}\in\mathbb{C}^{M_u\times N_u}$ is obtained as follows. An $M_uN_u$-length Zadoff--Chu (ZC) sequence is first arranged into an $M_u\times N_u$ matrix $\tilde{\boldX}_{\text{s},u}$, with entries
\begin{equation}
\tilde{\boldX}_{\text{s},u}[k,l]
=
e^{-j\pi\alpha \frac{(lM_u+k)(lM_u+k+1)}{M_uN_u}},
\end{equation}
$0 \leq k \leq M_u-1,
0 \leq l \leq N_u-1$,
where $\alpha$ is chosen to be co-prime to $M_uN_u$. An $N_u$-point unitary FFT is then applied along the rows of $\tilde{\boldX}_{\text{s},u}$ to obtain
$\boldX_{\text{s},u}$, as 
\begin{equation}
\boldX_{\text{s},u}[k,l]
=
\frac{1}{\sqrt{N_u}}
\sum_{j=0}^{N_u-1}
\tilde{\boldX}_{\text{s},u}[k,j]\,
e^{-j\frac{2\pi}{N_u}jl}.
\end{equation}
The transmit frame $\boldX_u\in\mathbb{C}^{M_u\times N_u}$ is then formed as 
\begin{equation}
\boldX_u = \boldX_{\text{s},u} + \boldX_{\text{d},u},
\end{equation}
where $\boldX_{\text{d},u}\in\alphabet^{M_u\times N_u}$ denotes the data symbol matrix and $\alphabet$ denotes the modulation alphabet. 

\begin{rem}
Two types of spread-pilots have been proposed for Zak-OTFS, namely, periodic twisted convolution based spread-pilot in \cite{isac} and ZC sequence based spread-pilot in \cite{zadoff_spread}. Both provide uniform energy distribution across the DD grid when $M_u$ and $N_u$ are co-prime. However, this co-prime condition cannot, in general, be satisfied in multiuser systems with heterogeneous frame sizes. In such cases, the ZC sequence based spread-pilot exhibits a lower PAPR than the periodic twisted convolution based spread-pilot, motivating its use in the proposed multiuser framework.
\end{rem}

\begin{figure}
    \centering
    \includegraphics[width=8.5cm, height=6.5cm]
       {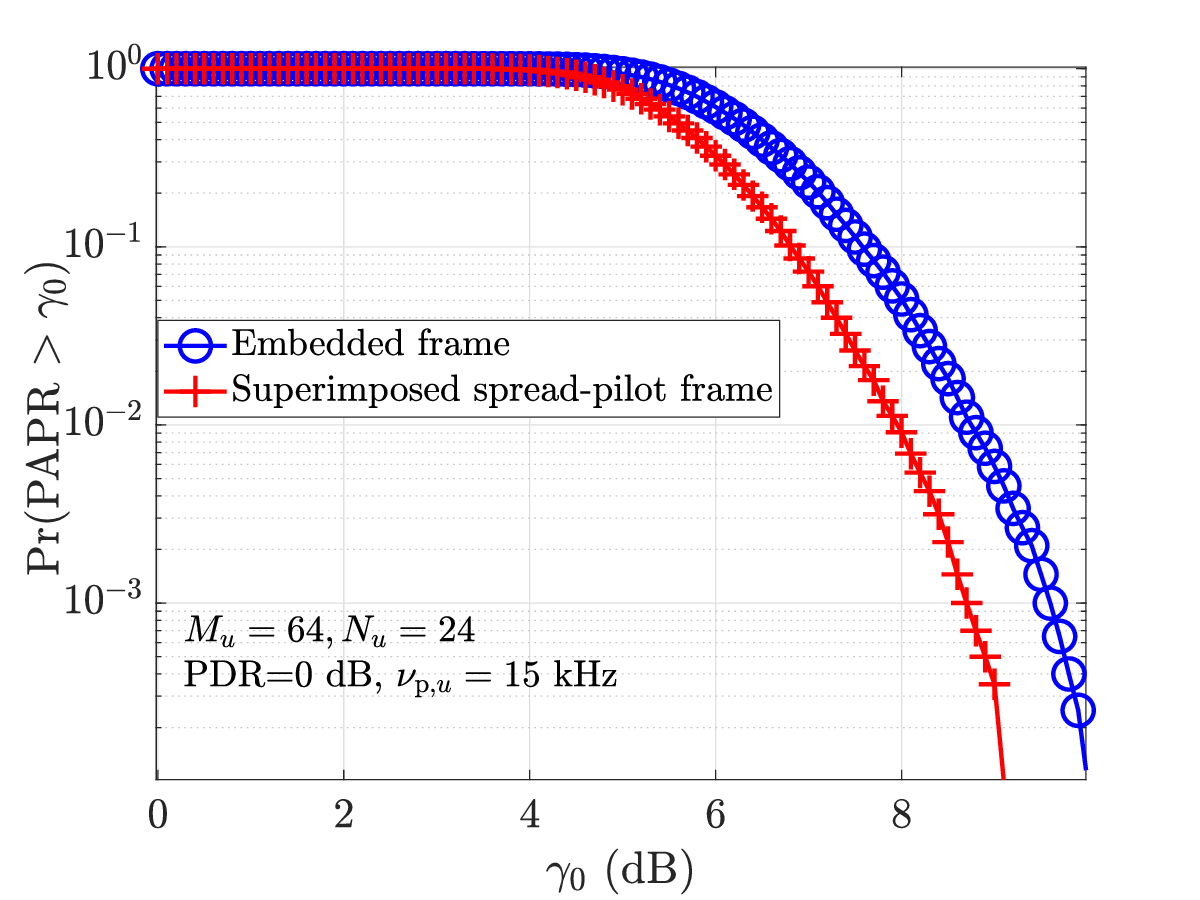}
    \vspace{-2mm}
    \caption{PAPR comparison of embedded pilot frame and superimposed spread-pilot frame for the $u$th user with $M_u = 64$, $N_u = 24$, $4$-QAM data symbols, and sinc filter.}
    \vspace{-2mm}
    \label{fig:papr}
\end{figure}

{\em CCDF of PAPR:} We note that the above superimposed spread-pilot structure, in addition to having the advantage of utilizing all $M_uN_u$ 
DD bins for data symbols, also yields a lower PAPR compared to the embedded pilot structure, in which a single pilot symbol is placed at the center of the DD frame, surrounded by a guard region of zeros (to mitigate pilot-data interference) with data symbols occupying the remaining DD bins. This is 
illustrated in Fig.~\ref{fig:papr}, which shows the CCDF of the PAPR for embedded pilot frame and superimposed spread-pilot frame for the $u$th user with $M_u = 64$, $N_u = 24$, $4$-QAM data symbols, and sinc filter.

\subsection{IOR estimation} 
\label{sec:est_mu}
To facilitate IOR estimation, we rewrite the system model in \eqref{eq:yu_final} in an equivalent form. We note that $\boldH_{u,v}$ is linear in the path gains $\{h_{i,v}\}_{i=1}^{P_v}$ of the $v$th user. Specifically, defining the $(u,v)$th \emph{path component matrix} $\boldG_{i,u,v} \triangleq
\boldH_{u,v}(1,\tau_{i,v},\nu_{i,v})$ as the IOR matrix corresponding to a channel of the $v$th user that contains only the $i$th path with unit gain, delay $\tau_{i,v}$, and Doppler $\nu_{i,v}$, the contribution of the $v$th user to the $u$th user's noise-free received signal can be written as
\begin{equation}
\boldy_{u,v}=
    \boldH_{u,v}\boldx_v
    =
    \underbrace{
    \big[
    \boldG_{1,u,v}\boldx_v\ \cdots\
    \boldG_{P_v,u,v}\boldx_v
    \big]
    }_{\triangleq\,\boldPhi_{u,v}}
    \boldh_v,
    \label{eq:mu_alternate_uv}
\end{equation}
where $\boldh_v = [h_{1,v}\ \cdots\ h_{P_v,v}]^T \in
\mathbb{C}^{P_v\times 1}$ denotes the path gain vector of the $v$th user. Substituting \eqref{eq:mu_alternate_uv} into \eqref{eq:yu_final}, the received signal vector $\boldy_u$ admits the alternate representation
\begin{align}
    \boldy_u
    &=
    \sum_{v=1}^{U}
    \boldPhi_{u,v}\boldh_v
    + \boldn_u
    \nonumber\\
    &=
    \underbrace{
    \big[
    \boldPhi_{u,1}\ \cdots\ \boldPhi_{u,U}
    \big]
    }_{\triangleq\,\boldPhi'}
    \underbrace{
    \begin{bmatrix}
    \boldh_1 \\ \vdots \\ \boldh_U
    \end{bmatrix}
    }_{\triangleq\,\boldh'}
    + \boldn_u,
    \label{eq:mu_alternate}
\end{align}
where $\boldPhi' \in \mathbb{C}^{M_uN_u \times
\sum_{v=1}^{U}P_v}$ and $\boldh' \in
\mathbb{C}^{\sum_{v=1}^{U}P_v \times 1}$ denote the
aggregate path component matrix and path gain vector, respectively. 
\begin{figure}
    \centering
    \includegraphics[width=8.5cm, height=6.5cm]
    {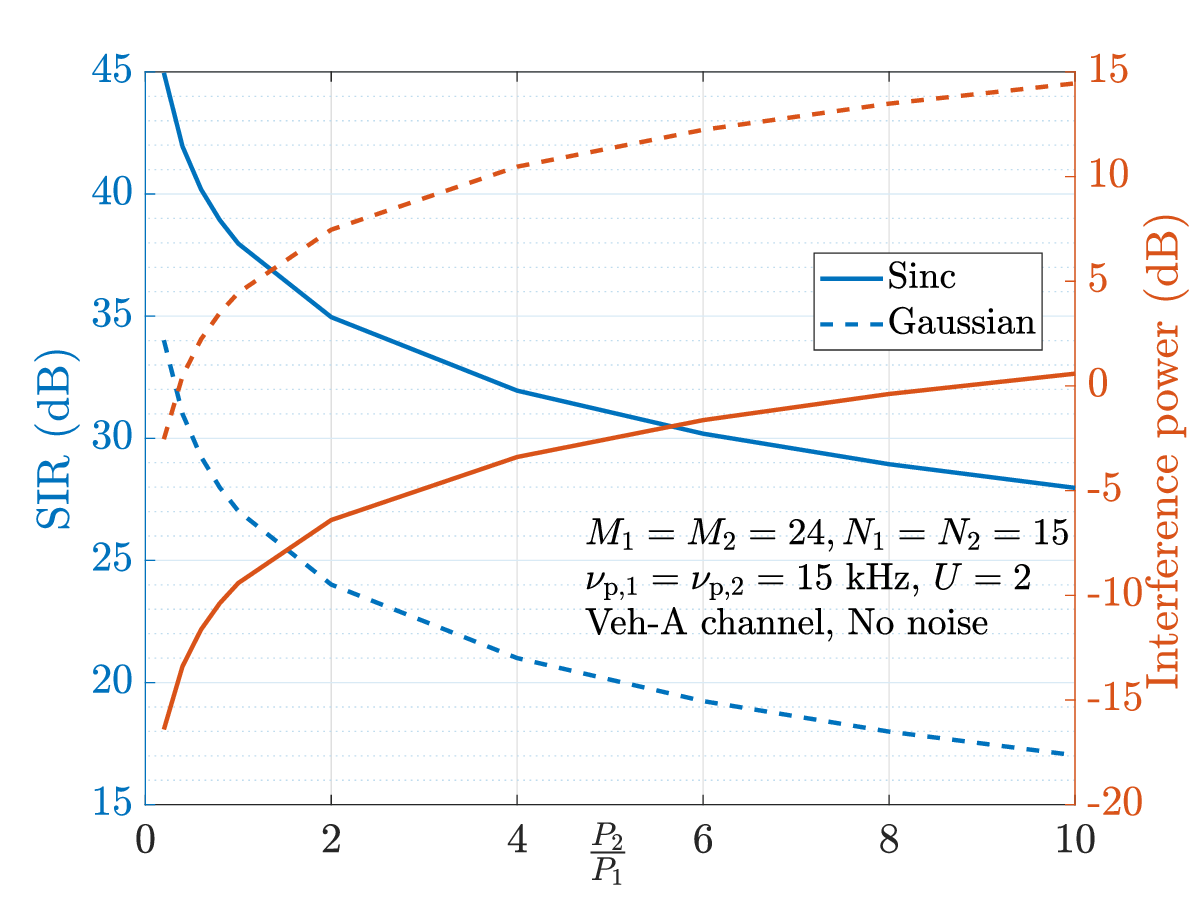}
    \vspace{-2mm}
    \caption{SIR and interference power at User~1's receiver as a function of inter-user power ratio $P_2/P_1$ for a two-user 
    system with $M_1=M_2=24$, $N_1=N_2=15$, $\nu_{\text{p}}=15$~kHz, Veh-A channel, no noise, and superimposed spread-pilot.}
    \label{fig:two_user_SIR}
    \vspace{-2mm}
\end{figure}

We note that \eqref{eq:mu_alternate} serves as the key structure exploited in the dictionary-based IOR estimation scheme. A critical observation that significantly simplifies the estimation problem is that the IUI terms in \eqref{eq:mu_alternate}, i.e., $\sum_{v\neq u}\boldPhi_{u,v}\boldh_v$, are negligible in the Zak-OTFS multiuser setting considered here. This is because the TF-shift-based multiple access scheme places distinct users in non-overlapping regions of the TF plane (Sec.~\ref{sec:sys_model}), so that the effective channel $h_{\eff,u,v}(\tau,\nu)$ for $v\neq u$ is strongly attenuated. To validate this, in Fig.~\ref{fig:two_user_SIR}, we consider a two-user system with superimposed spread-pilot and plot the signal-to-interference power ratio (SIR) and interference power at User~1's receiver as a function of the inter-user power ratio $P_2/P_1$, where $P_1$ and $P_2$ denote the transmit powers of User~1 and User~2, respectively, for  sinc and Gaussian filters with $M_1=M_2=24$, $N_1=N_2=15$,
$\nu_{\text{p},1}=\nu_{\text{p},2}=15$~kHz, Veh-A channel, and no noise. It is observed that even when User~2 transmits at ten times the power of User~1 (i.e., $P_2/P_1=10$), the SIR at User~1's receiver remains above $15$~dB for the Gaussian pulse and above $28$~dB for the sinc pulse, confirming that the IUI very small in both cases. Consequently, \eqref{eq:mu_alternate} simplifies to
\begin{eqnarray}
    \boldy_u
    & \hspace{-2mm} \approx & \hspace{-2mm}
    \boldPhi_{u,u}\boldh_u
    + \boldn_u \nonumber \\
    & \hspace{-2mm} = & \hspace{-2mm}
    \underbrace{
    \big[
    \boldG_{1,u}\boldx_u\ \cdots\
    \boldG_{P_u,u}\boldx_u
    \big]
    }_{\triangleq\,\boldPhi_u}
    \boldh_u
    + \boldn_u,
    \label{eq:mu_alternate_simplified}
\end{eqnarray}
where we have used $\boldPhi_{u}\triangleq\boldPhi_{u,u},\boldG_{i,u}\triangleq\boldG_{i,u,u}$. We note that \eqref{eq:mu_alternate_simplified} has the same form as the single-user alternate system model, with the $u$th user's own IOR matrix $\boldH_{u,u}$ playing the role of $\boldH$. This decoupling across users implies that the IOR of each user can be estimated independently at the BS, using only that user's transmitted frame, without requiring
knowledge of the other users' signals.

\subsubsection{DD dictionary and path component matrix}
The physical channel of the $u$th user is approximated using a DD dictionary $\mathcal{D}_u =
\tilde{\boldtau}_u \times \tilde{\boldnu}_u$, with
\begin{equation}
    \tilde{\boldtau}_u
    =
    \left\{
        \frac{k\tau_{\text{p},u}}{M_us_{\tau,u}}
        \;\middle|\;
        k = 0, 1, \ldots, k_{\max,u}s_{\tau,u}
    \right\},
\end{equation}
\begin{equation}
    \tilde{\boldnu}_u
    =
    \left\{
        \frac{l\nu_{\text{p},u}}{N_us_{\nu,u}}
        \;\middle|\;
        l = -l_{\max,u}s_{\nu,u}, \ldots,
        l_{\max,u}s_{\nu,u}
    \right\},
\end{equation}
where $k_{\max,u} =
\lceil\tau_{\max,u}M_u/\tau_{\text{p},u}\rceil$,
$l_{\max,u} = \lceil\nu_{\max,u}N_u/\nu_{\text{p},u}\rceil$,
and $s_{\tau,u},s_{\nu,u}\in\mathbb{R}^+\backslash\{0\}$ are the delay and Doppler discretization parameters of the $u$th user, respectively. The corresponding path component matrices are generated as $\tilde{\boldG}_{i,u} =
\boldH_{u,u}(1,\tilde{\tau}_{i,u},\tilde{\nu}_{i,u})$
for each $(\tilde{\tau}_{i,u},\tilde{\nu}_{i,u})\in\mathcal{D}_u$. The pilot and data contributions to the measurement matrix of the $u$th user at the BS are
\begin{align}
    \tilde{\boldPhi}_{\text{s},u}
    &=
    \big[
    \tilde{\boldG}_{1,u}\boldx_{\text{s},u}
    \ \cdots\
    \tilde{\boldG}_{|\mathcal{D}_u|,u}
    \boldx_{\text{s},u}
    \big],
    \label{eq:Phi_p_u}
    \\
    \tilde{\boldPhi}_{\text{d},u}
    &=
    \big[
    \tilde{\boldG}_{1,u}\boldx_{\text{d},u}
    \ \cdots\
    \tilde{\boldG}_{|\mathcal{D}_u|,u}
    \boldx_{\text{d},u}
    \big],
    \label{eq:Phi_d_u}
\end{align}
so that the aggregate path component matrix is
\begin{equation}
    \tilde{\boldPhi}_u
    =
    \tilde{\boldPhi}_{\text{s},u}
    +
    \tilde{\boldPhi}_{\text{d},u}.
    \label{eq:Phi_u}
\end{equation}

\subsubsection{Iterative IOR estimation algorithm}
Using \eqref{eq:mu_alternate_simplified} and
\eqref{eq:Phi_u}, the received signal at the BS for the $u$th user is written as
\begin{equation}
    \boldy_u
    \approx
    \tilde{\boldPhi}_u\,\tilde{\boldh}_u
    + \boldn_u,
    \label{eq:mu_y}
\end{equation}
where $\tilde{\boldh}_u \in
\mathbb{C}^{|\mathcal{D}_u| \times 1}$ denotes the
dictionary path gain vector of the $u$th user.
The maximum-likelihood estimate of $\tilde{\boldh}_u$ is given by
\begin{align}
    \widehat{\tilde{\boldh}}_u^{(t)}
    &=
    \left(
    \Big(\tilde{\boldPhi}_u^{(t)}\Big)^H
    \boldR_u^{-1}
    \tilde{\boldPhi}_u^{(t)}
    \right)^{-1}
    \Big(\tilde{\boldPhi}_u^{(t)}\Big)^H
    \boldR_u^{-1}
    \boldy_u,
    \label{eq:mu_h_hat}
\end{align}
where $\boldR_u$ denotes the noise covariance matrix
at the BS for the $u$th user. Since
$\tilde{\boldPhi}_{\text{d},u}$ depends on the unknown data symbols of the $u$th user, a direct computation of \eqref{eq:mu_h_hat} is not possible. The IOR of each user is therefore estimated independently using the iterative algorithm summarized in \textbf{Algorithm~\ref{alg:mu_ior}}, which is run separately for each $u \in \{1,\ldots,U\}$.

The algorithm is initialized with the data vector set to zero. In the $t$th iteration, the data decision from the $(t\!-\!1)$th iteration is used to update $\tilde{\boldPhi}_{\text{d},u}^{(t)}$ (step~4). The updated aggregate path component matrix $\tilde{\boldPhi}_u^{(t)}$ is then formed (step~5), and the path gain vector is re-estimated using \eqref{eq:mu_h_hat} (step~6). The estimated
coefficients are used to reconstruct the IOR matrix
$\hat{\boldH}_{u,u}^{(t)}$ (step~6), from which an
updated data decision is obtained on the residual
signal (step~7). This process continues until
$T_{\max}$ iterations are reached or the path gain
estimate converges.

\begin{algorithm}[t]
\caption{Proposed IOR estimation algorithm
(run independently for each user $u \in \{1,\ldots,U\}$)}
\label{alg:mu_ior}
\begin{algorithmic}[1]
\STATE \textbf{Input:}
$\{\tilde{\boldG}_{i,u}\}_{i=1}^{|\mathcal{D}_u|}$,
$\tilde{\boldPhi}_{\text{p},u}$,
$\boldy_u$,
$\boldR_u^{-1}$,
$T_{\max}$
\STATE \textbf{Initialize:}
$\hat{\boldx}_{\text{d},u}^{(0)}
= \mathbf{0}_{M_uN_u\times 1}$,\quad
$t \leftarrow 1$
\REPEAT
\STATE
$\tilde{\boldPhi}_{\text{d},u}^{(t)}
\leftarrow
\big[
\tilde{\boldG}_{1,u}
\hat{\boldx}_{\text{d},u}^{(t-1)}
\ \cdots\
\tilde{\boldG}_{|\mathcal{D}_u|,u}
\hat{\boldx}_{\text{d},u}^{(t-1)}
\big]$
\STATE
$\tilde{\boldPhi}_u^{(t)}
\leftarrow
\tilde{\boldPhi}_{\text{s},u}
+
\tilde{\boldPhi}_{\text{d},u}^{(t)}$
\STATE
$\widehat{\tilde{\boldh}}_u^{(t)}
\leftarrow
\left(
\big(\tilde{\boldPhi}_u^{(t)}\big)^H
\boldR_u^{-1}
\tilde{\boldPhi}_u^{(t)}
\right)^{-1}
\big(\tilde{\boldPhi}_u^{(t)}\big)^H
\boldR_u^{-1}
\boldy_u$
\STATE
$\hat{\boldH}_{u,u}^{(t)}
\leftarrow
\displaystyle\sum_{i=1}^{|\mathcal{D}_u|}
\widehat{\tilde{h}}_u^{(t)}[i]\,
\tilde{\boldG}_{i,u}$
\STATE Detect $\hat{\boldx}_{\text{d},u}^{(t)}$
from residual
$\boldy_u - \hat{\boldH}_{u,u}^{(t)}
\boldx_{\text{s},u}$
using $\hat{\boldH}_{u,u}^{(t)}$
\STATE $t \leftarrow t+1$
\UNTIL{$t > T_{\max}$ \textbf{or}
$\|\widehat{\tilde{\boldh}}_u^{(t)}
-\widehat{\tilde{\boldh}}_u^{(t-1)}\| < \eta$}
\STATE \textbf{Output:}
$\hat{\boldH}_{u,u}
\leftarrow \hat{\boldH}_{u,u}^{(t)}$
\end{algorithmic}
\end{algorithm}

\begin{rem}
The decoupling of users established in
\eqref{eq:mu_alternate_simplified} implies that
\textbf{Algorithm~\ref{alg:mu_ior}} scales with the number of users $U$, since the computational
complexity of IOR estimation per user is identical to the single user case and the $U$ instances of the
algorithm can be run in parallel at the BS.
\end{rem}

\section{Results and Discussions} 
\label{sec:results}

\begin{figure}[t]
    \centering
    \includegraphics[width=0.85\linewidth]{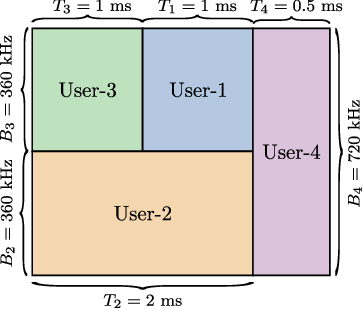}
    \caption{TF plane allocation for the four-user  
    system, illustrating the non-overlapping TF
    support regions of the four users achieved via TF shifts.}
    \label{fig:mu_tf}
\end{figure}

\begin{table}[t]
\caption{Simulation parameters of the four-user system.} 
\centering
\begin{tabular}{|c||c|c|c|c|}
\hline
Parameters & User~1 & User~2 & User~3 & User~4 \\ \hline \hline
$M_u$                    & 24   & 24  & 12     & 24   \\ \hline
$N_u$                    & 15   & 30  & 30     & 15   \\ \hline
$\nu_{\text{p},u}$ (kHz) & 15   & 15  & 30     & 30   \\ \hline
$B_u$ (kHz)              & 360  & 360 & 360    & 720  \\ \hline
$T_u$ (ms)               & 1    & 2   & 1      & 0.5  \\ \hline
$\tau_u$ (ms)            & 0.5  & 0   & $-$0.5 & 1.25 \\ \hline
$\nu_u$ (kHz)            & 360  & 0   & 360    & 0    \\ \hline
\end{tabular}
\label{tab:sim_params}
\end{table}

In this section, we present simulation results for a four-user Zak-OTFS system. The simulation parameters are summarized in Table~\ref{tab:sim_params} and the corresponding TF plane allocation is illustrated in Fig.~\ref{fig:mu_tf}. The four users are assigned
heterogeneous DD grid sizes and TF shifts
$(\tau_u,\nu_u)$ such that their TF support regions
are non-overlapping. We note that the users differ
not only in their delay and Doppler periods
$(\tau_{\text{p},u},\nu_{\text{p},u})$ but also in
their bandwidths $B_u$ and frame durations $T_u$,
thereby representing a heterogeneous multiuser scenario. The doubly-dispersive channel of each user is modeled using the Veh-A power delay profile (PDP) with $P_u=6$ paths \cite{VehA}, where the Doppler shift of the $i$th path of the $u$th user is generated as $\nu_{i,u}=\nu_{\max}\cos(\theta_{i,u})$,
with $\theta_{i,u}$ drawn independently and uniformly from $[-\pi,\pi)$ and $\nu_{\max}=815$~Hz. The superimposed spread-pilot frame described in Sec.~\ref{sec:spread_pilot} is used for all users, with the ZC sequence root set to $\alpha=7$. The IOR estimation algorithm
(\textbf{Algorithm~\ref{alg:mu_ior}}) is run with
a maximum of $T_{\max}=15$ iterations, convergence
threshold $\eta=10^{-3}$, and discretization
parameters $s_{\tau,u}=s_{\nu,u}=2$. Data detection is carried out using the least-squares minimum-residual based interference cancellation (LSMR-IC) detector \cite{lsmr}, and all simulations use $4$-QAM data symbols. The pilot-to-data power ratio (PDR), defined as $\text{PDR}\triangleq E_{\text{p}}/E_{\text{d}}$, where $E_{\text{p}}$ and $E_{\text{d}}$ denote the pilot and data energies in the transmitted frame, respectively, is fixed at $0$~dB throughout. Data SNR (DSNR) is defined as $\frac{E_d}{N_0MN}$.

Performance is evaluated in terms of three metrics.
The first is the normalized mean square error (NMSE) of the estimated IOR matrix of the $u$th user, defined as
$\mathrm{NMSE}_u
    \triangleq
    \mathbb{E}\!\left[
    \frac{
    \left\lVert\boldH_{u,u}-\hat{\boldH}_{u,u}
    \right\rVert_F^2
    }{
    \left\lVert\boldH_{u,u}\right\rVert_F^2
    }
\right].$
The second is the uncoded bit error rate (BER), and the third is the spetral efficiency (SE) of the $u$th user, defined as \cite{SE}
\begin{equation}    
\mathrm{SE}_u =
\frac{R_{\mathrm{c}}\,\beta_u\,\log_2|\alphabet|}{B_uT_u} \left(1-\mathrm{BLER}\right),
\label{eq:se}
\end{equation}
where $R_{\mathrm{c}}=1/2$ denotes the coding rate
of the convolutional code employed, $\beta_u=M_uN_u$ denotes the number of DD bins carrying information symbols in the superimposed spread-pilot frame, and $\mathrm{BLER}$ denotes the block error rate. The SE metric is computed using coded symbols, whereas the BER and NMSE metrics are
evaluated using uncoded data symbols. Since User~1
experiences interference from all three remaining
users, it represents the most vulnerable user in the considered four-user setup, and therefore all results are presented for User~1.

\subsection{IOR estimation performance}
\begin{figure}[t]
    \centering
    \includegraphics[width=8.5cm, height=6.5cm]
    {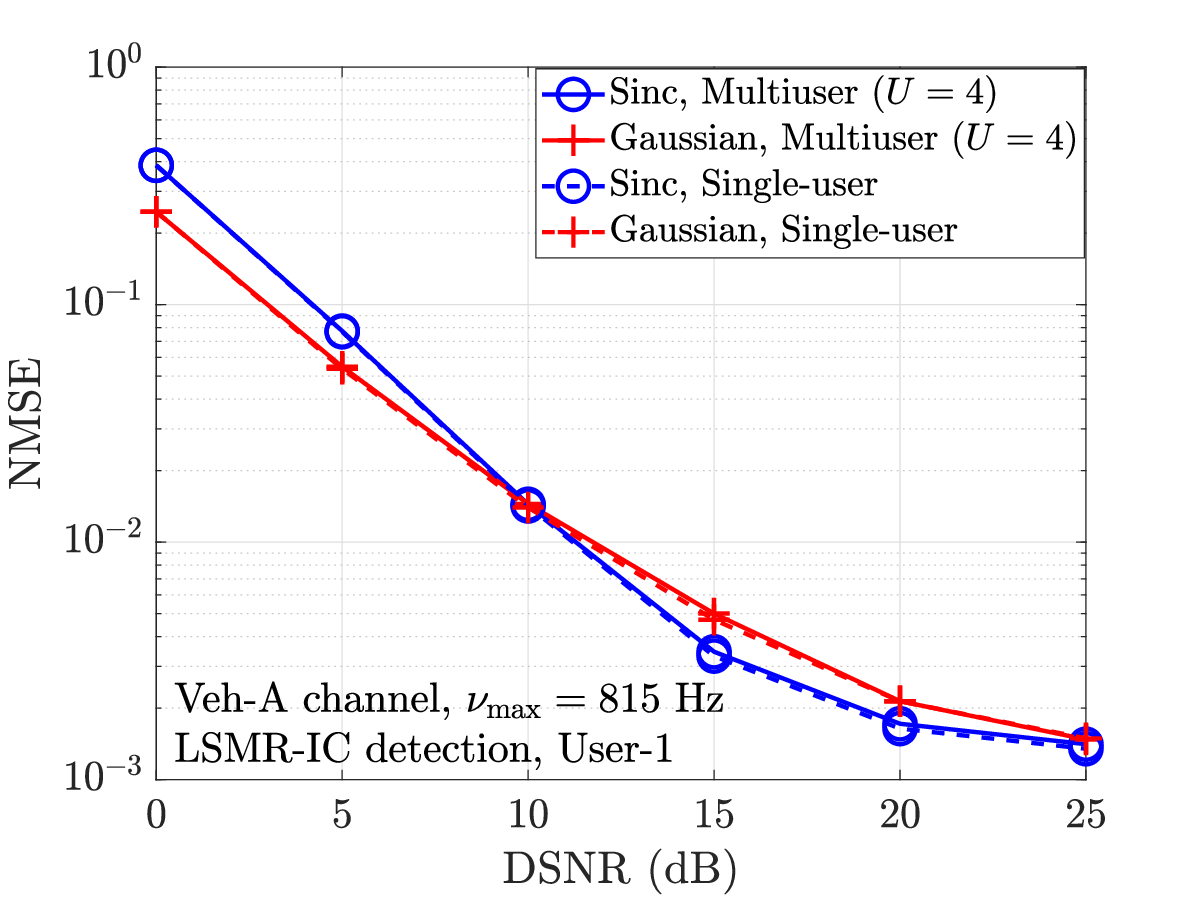}
    \vspace{-2mm}
    \caption{NMSE of the estimated IOR matrix of User~1 versus DSNR in four-user and single-user systems with sinc and Gaussian filters.} 
    \vspace{-2mm}
    \label{fig:nmse}
\end{figure}

In Fig.~\ref{fig:nmse}, we plot the NMSE performance of User~1 as a function of DSNR for 
sinc and Gaussian filters. Single-user performance is also shown for comparison. It can be observed that the NMSE performance for four users and single user are nearly the same for both filters. This is consistent with the observation in Sec.~\ref{sec:est_mu} that the IUI is very small 
(Fig.~\ref{fig:two_user_SIR}). This shows that, with the TF-shift-based multiple access, the IOR of each user can be estimated (without much loss in estimation accuracy) as if the other users were absent.

\subsection{Uncoded BER performance}
\begin{figure}[t]
    \centering
    \includegraphics[width=8.5cm, height=6.5cm]
    {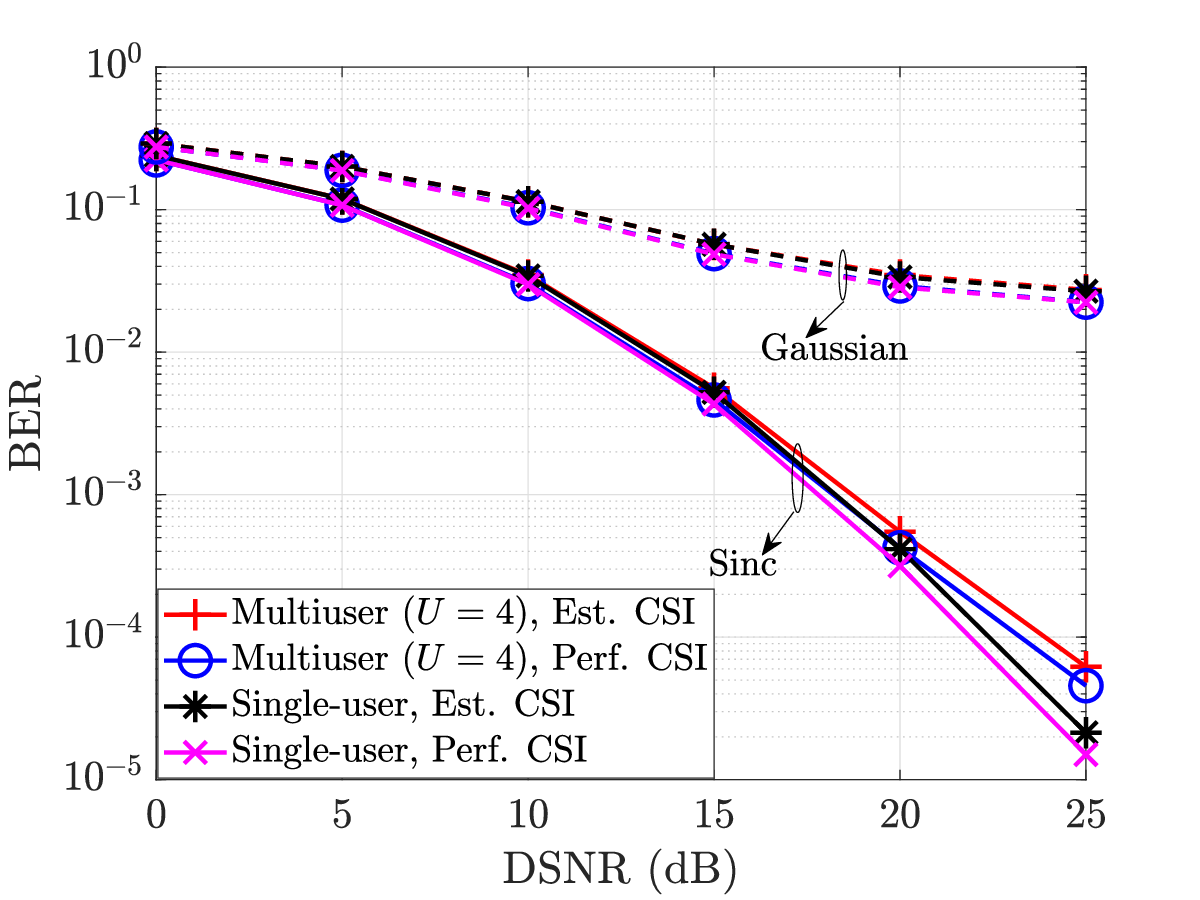}
    \vspace{-2mm}
    \caption{Uncoded BER of User~1 versus DSNR in four-user and single-user systems 
    with sinc and Gaussian filters.}
    \label{fig:ber}
    \vspace{-2mm}
\end{figure}

In Fig.~\ref{fig:ber}, we plot the uncoded BER of User~1 as a function of DSNR for sinc and Gaussian filters with estimated CSI (using \textbf{Algorithm~\ref{alg:mu_ior}}). Performance with perfect CSI is also shown for comparison.
The Gaussian filter is found to yield significantly poorer BER performance compared to the sinc filter for both perfect CSI and estimated CSI. This is attributed to the higher interference power induced by the Gaussian filter on adjacent DD symbols, resulting in a residual interference floor and consequent flooring at a higher BER. In contrast, the sinc filter causes significantly less interference to adjacent DD symbols due to its pulse shape (nulls at sampling points), leading to a better BER. We further see that the presence of other users does not significantly degrade the BER of User 1 (because of low IUI) and that BER performance close to that with perfect CSI is achieved with estimated CSI.   

\begin{figure}
    \centering
    \includegraphics[width=8.5cm, height=6.5cm]
      {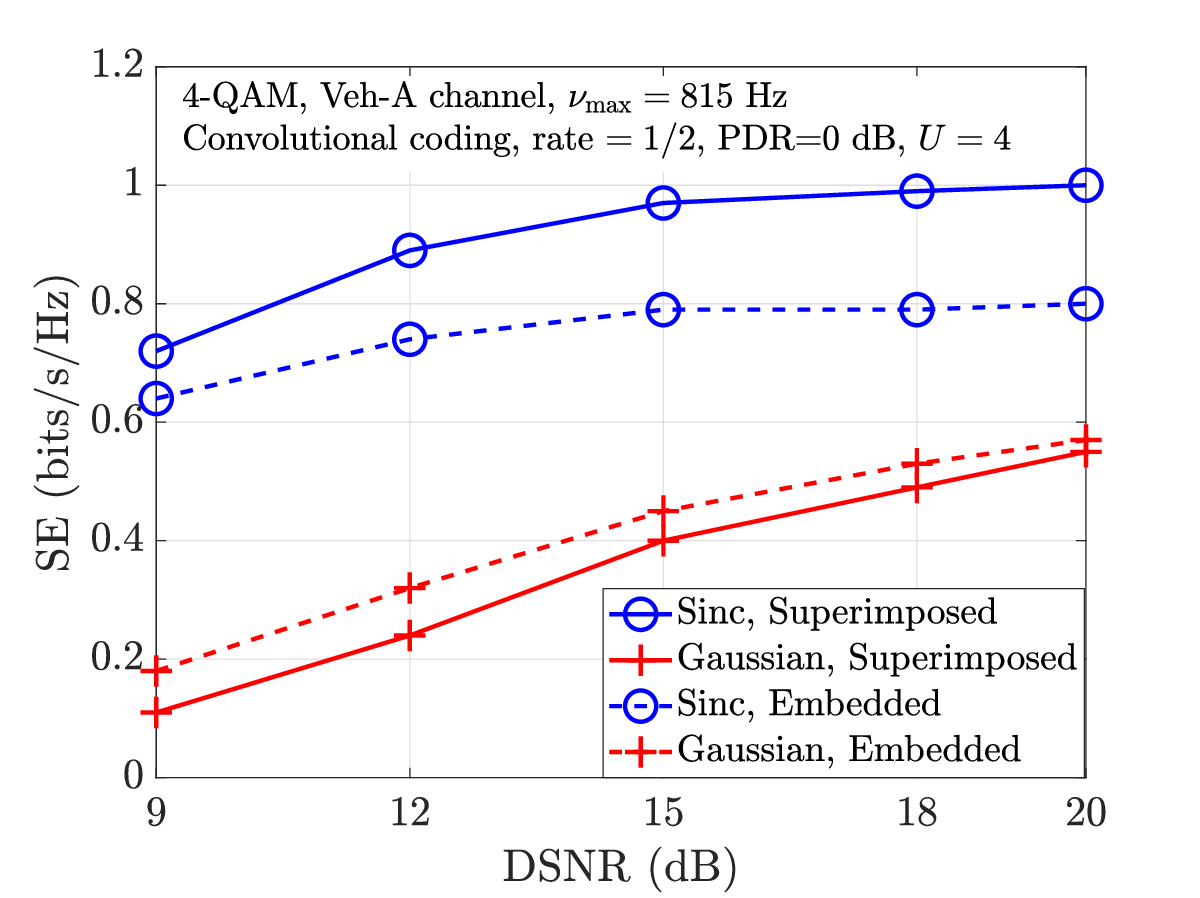}
    \vspace{-2mm}
    \caption{SE of User~1 versus DSNR in a four-user system with sinc and Gaussian filters 
    for superimposed spread-pilot and embedded pilot frames.}
    \label{fig:se}
    \vspace{-2mm}
\end{figure}

\begin{figure*}[t]
    \centering
    \subfigure[BLER vs $P_2/P_1$]
    {\label{fig:toy_bler}
    \includegraphics[width=0.475\linewidth]{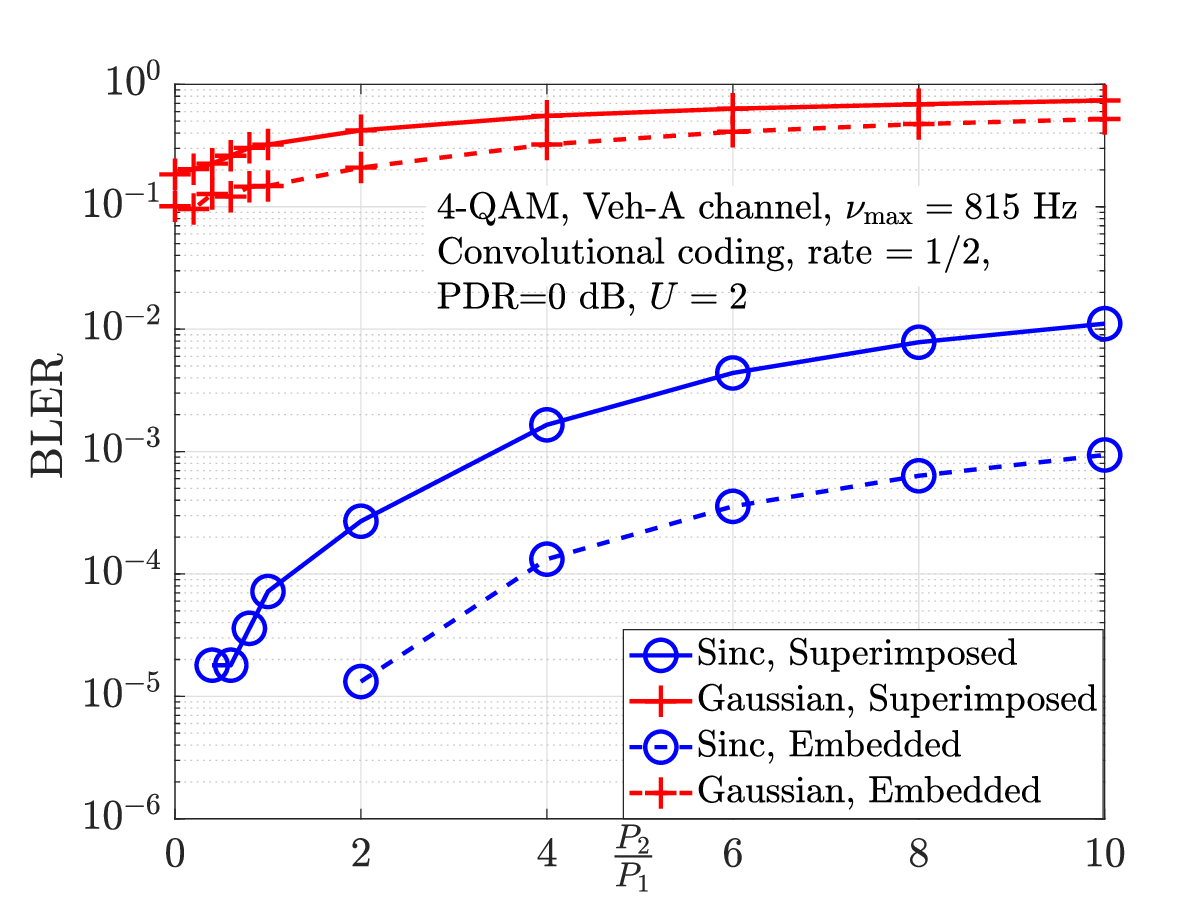}}
    \hspace{-4mm}
    \subfigure[SE vs $P_2/P_1$]
    {\label{fig:toy_se}
    \includegraphics[width=0.475\linewidth]{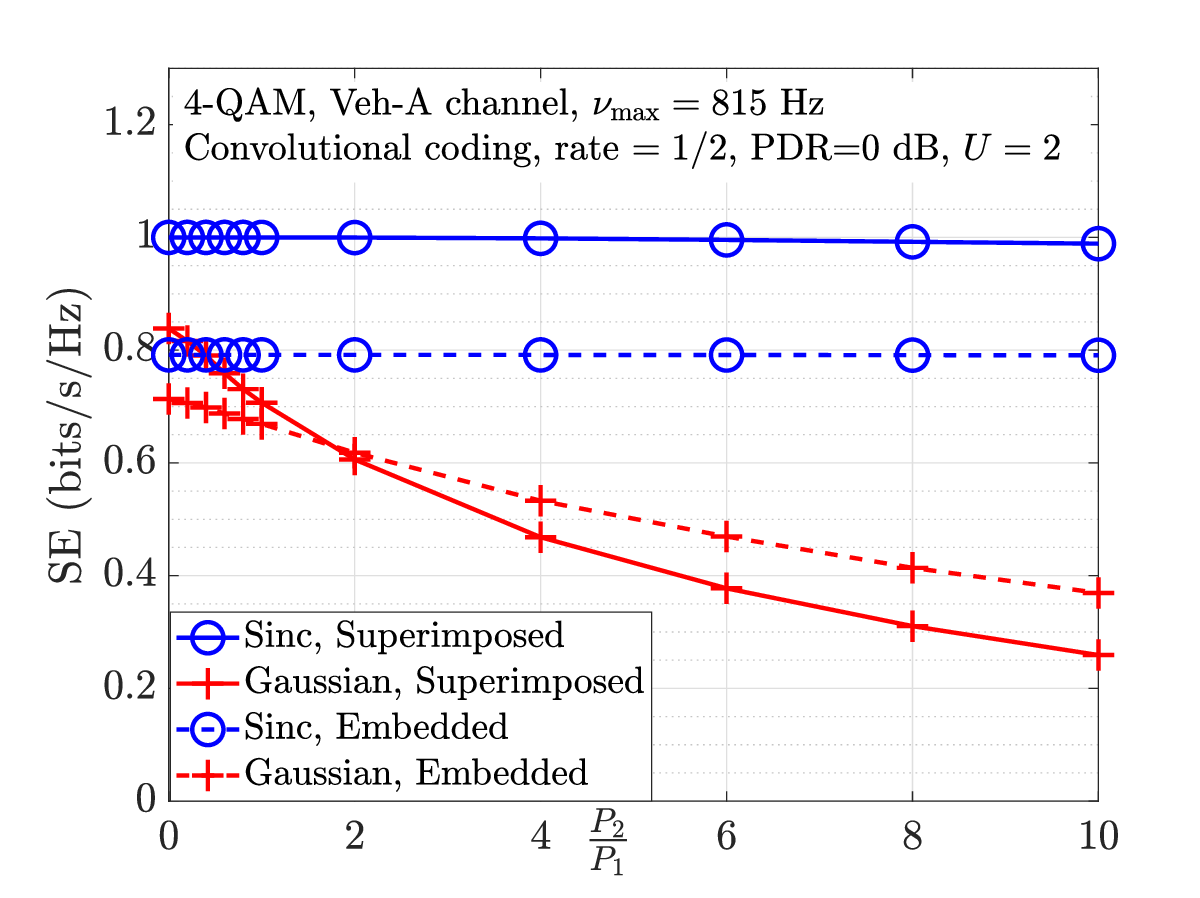}}
    \caption{
    BLER and SE of User~1 as a function
    of the inter-user power ratio for a two-user
    uplink Zak-OTFS system with $M_1=M_2=24$,
    $N_1=N_2=15$ for superimposed spread-pilot and embedded pilot frames.}
    \label{fig:toy_example}
    \vspace{-2mm}
\end{figure*}

\subsection{Spectral efficiency performance}
In Fig.~\ref{fig:se}, we compare the SE achieved by User 1 with superimposed spread-pilot and embedded pilot frames. We see that, with sinc filter, the superimposed frame achieves higher SE compared to that with the embedded frame. This is because all the $M_uN_u$ DD bins in a superimposed spread-pilot frame carry data symbols, whereas some DD bins in an embedded pilot frame are earmarked as pilot and guard bins where data symbols are not sent, thus reducing the achievable SE\footnote{In an embedded pilot frame, a single pilot symbol is placed at the center of the $M_u\times N_u$ DD grid. For a guard region of width $2k_{\max}$ along the delay dimension (above and below the pilot location) and spanning the entire Doppler dimension left vacant to mitigate pilot-data interference, only $M_uN_u - (2k_{\max}+1)N_u = 285$ out of $M_uN_u=360$ DD bins are available for data symbols.}. For the Gaussian filter, however, the picture reverses; embedded frame achieves higher SE compared to superimposed
frame. This reversal is not immediately explained by the NMSE or BER results alone, and warrants a more careful analysis, which we carry out next.

\subsubsection*{Analysis of the SE behavior in Fig.~\ref{fig:se}}
To explain the better SE performance of Gaussian filter in embedded pilot frame compared to superimposed spread-pilot frame in Fig.~\ref{fig:se}, we consider a two-user noise-free setup. In Figs.~\ref{fig:toy_bler} and~\ref{fig:toy_se}, we plot the BLER and SE of User~1, respectively, as functions of the inter-user power ratio $P_2/P_1$. 

The impact of the filter on detection performance in terms of BLER is reflected in Fig.~\ref{fig:toy_bler}. Consistent with the BER
results in Fig.~\ref{fig:ber}, the sinc filter
achieves significantly lower BLER compared to the Gaussian filter for both embedded and superimposed pilot frames. Also, for a given filter, the embedded frame consistently achieves a lower BLER than the superimposed frame because of the pilot-data interference inherent in the superimposed pilot
structure.

The corresponding impact on the SE is shown in
Fig.~\ref{fig:toy_se}. For the sinc filter, the BLER
remains sufficiently low for both pilot structures
such that the $(1-\mathrm{BLER})$ term in
\eqref{eq:se} is close to unity. Consequently, the
SE is primarily determined by the number of
data-carrying DD bins, and the superimposed frame
consistently achieves a higher SE by utilizing all
$M_uN_u$ DD bins for data transmission.

For the Gaussian filter, although the superimposed
frame has a higher BLER than the embedded frame, the
resulting BLER penalty at low values of $P_2/P_1$ is
insufficient to offset the advantage of utilizing all $M_uN_u$ data-carrying DD bins. Consequently, the superimposed frame achieves a higher SE than the
embedded frame. As $P_2/P_1$ increases, however, the
combined effect of increasing IUI and pilot-data interference causes the BLER of the superimposed frame to increase more rapidly than that of the embedded frame. Beyond a certain inter-user power ratio, the resulting reduction in the $(1-\mathrm{BLER})$ term in \eqref{eq:se} outweighs
the benefit of the additional data-carrying DD bins.
Consequently, the SE of the Gaussian superimposed
frame falls below that of the Gaussian embedded
frame, explaining the SE reversal observed in
Fig.~\ref{fig:toy_se}. The same behavior is reflected in the four-user scenario of Fig.~\ref{fig:se}, where the aggregate interference from the remaining users places the operating point in this high interference regime.

\subsection{Robustness to different channel models/max. Dopplers} 
\begin{figure}[t]
    \centering
    \includegraphics[width=8.5cm, height=6.5cm]
    {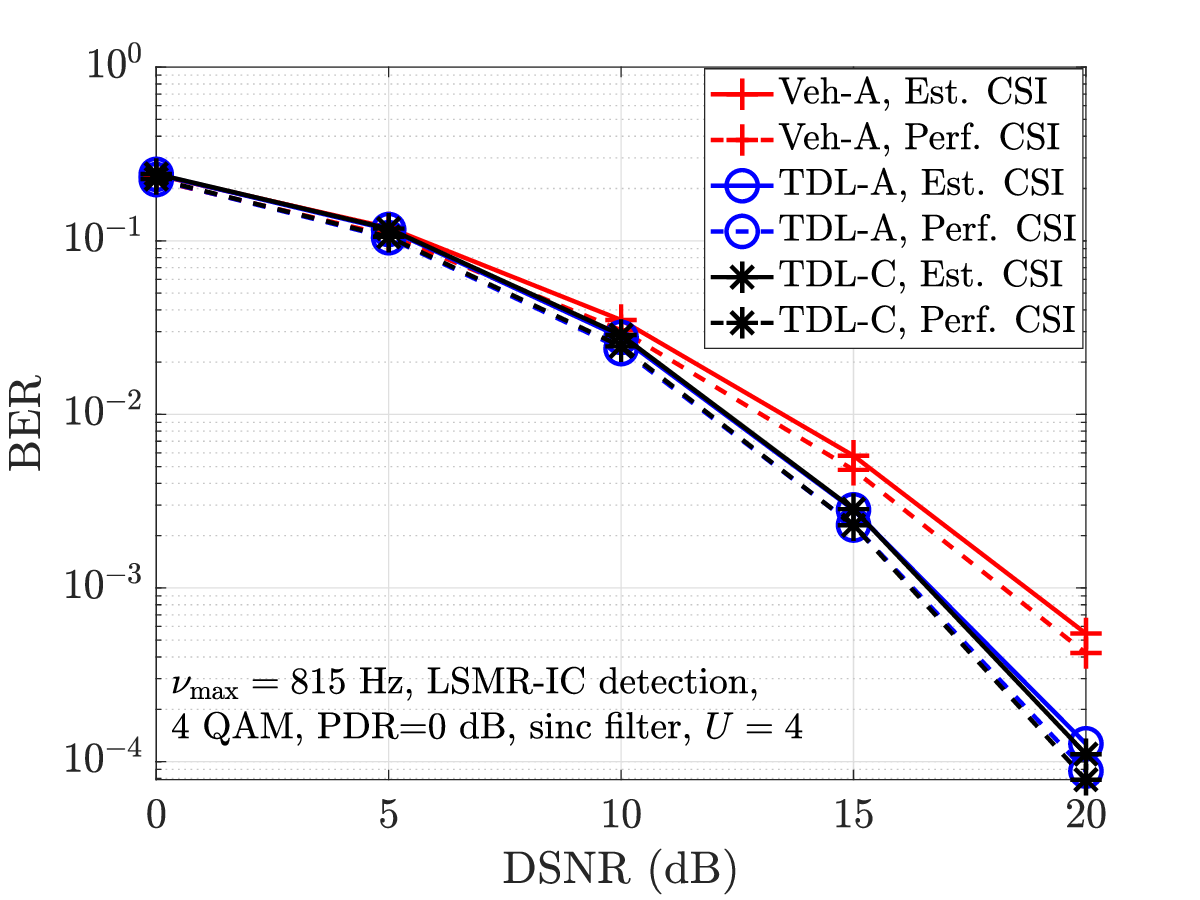}
    \vspace{-2mm}
    \caption{Uncoded BER of User 1 in a four-user system for Veh-A, TDL-A, TDL-C channel models with sinc filter and superimposed spread-pilot.}
    \label{fig:chlmodels}
    \vspace{-2mm}
\end{figure}

\begin{figure}
    \centering
    \includegraphics[width=8.5cm, height=6.5cm]
    {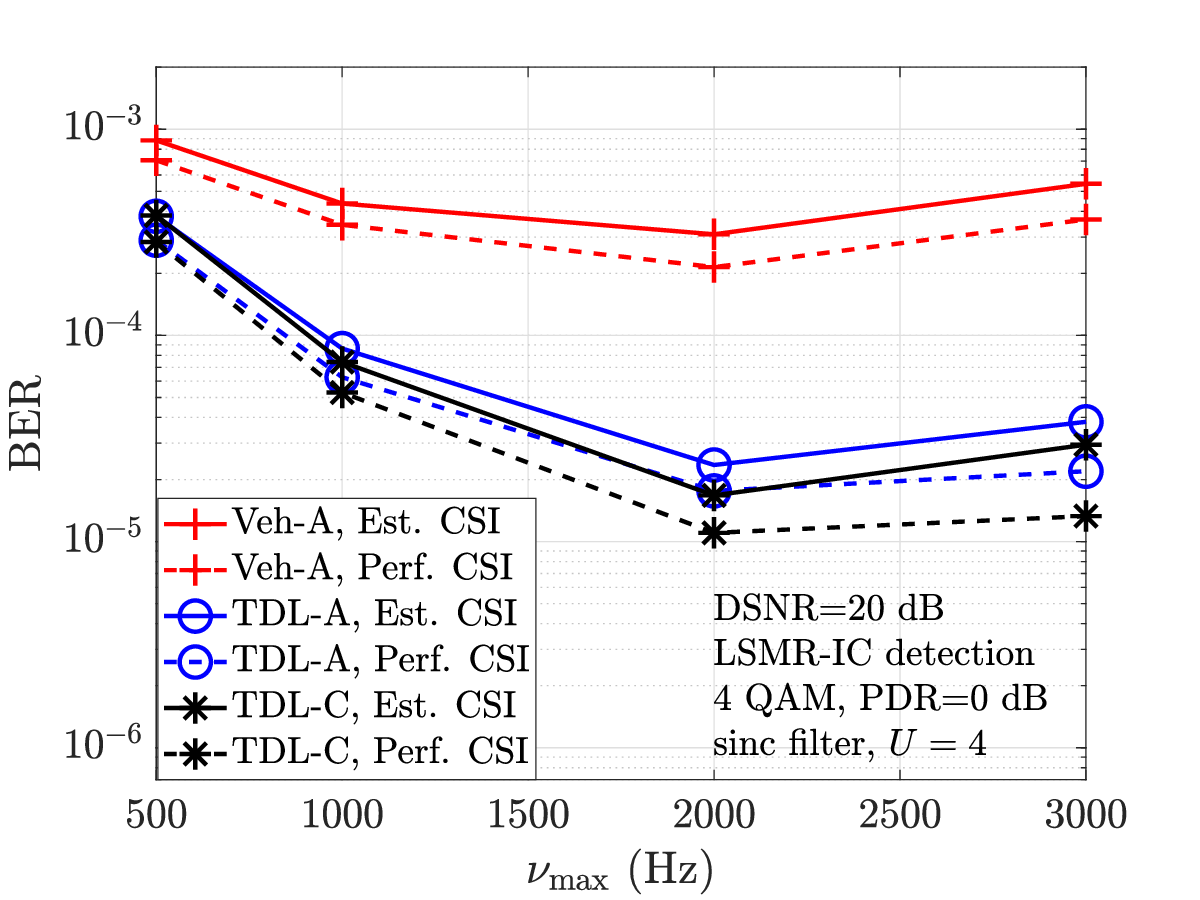}
    \vspace{-2mm}
    \caption{Uncoded BER of User 1 as a function of maximum Doppler in a four-user system with sinc filter and superimposed spread-pilot.}
    \label{fig:numax}
    \vspace{-2mm}
\end{figure}

Having established the performance under the Veh-A channel model at $\nu_{\max}=815$~Hz, we now examine the robustness to variations in the channel PDP and maximum Doppler. All results in this subsection are for sinc pulse with superimposed spread pilot. We note that TDL-A and TDL-C are rich scattering channel models with as many as $23$ and $24$ propagation paths, respectively, compared to the $6$-path Veh-A model.

In Fig.~\ref{fig:chlmodels}, we plot the uncoded BER of User~1 as a function of DSNR under the three channel models, with estimated CSI and perfect CSI. It is observed that the BER curves of TDL-A and TDL-C are nearly identical, and both achieve a lower BER than Veh-A at moderate-to-high DSNRs. The lower BER of TDL-A and TDL-C relative to Veh-A is attributable to the larger number of propagation paths in TDL-A and TDL-C, which provides greater multipath diversity and thus better BER performance\footnote{{In \cite{Fathima_VTC}, it has been shown that Zak-OTFS with $P$ paths achieves full diversity of $P$ under maximum-likelihood detection. Consequently, TDL-A and TDL-C with a larger number of paths than Veh-A achieve better performance compared to Veh-A, as observed in Figs. (\ref{fig:chlmodels}) and (\ref{fig:numax}).}}. For all the three channel models, the gap between the estimated and perfect CSI curves is small, demonstrating the effectiveness of the IOR estimation algorithm for different channel models.

In Fig.~\ref{fig:numax}, we plot the uncoded BER of User~1 as a function of $\nu_{\max}$ at DSNR~$=20$~dB, for the three channel models with estimated CSI and perfect CSI. It is observed that for all the three channel models, the BER initially decreases as $\nu_{\max}$ increases from $500$~Hz to $2000$~Hz, and then increases slightly as $\nu_{\max}$ increases further to $3000$~Hz. The initial decrease occurs due to reduced inter-path interference. With randomly chosen Dopplers of different paths (from uniform distribution) in the range $-\nu_{\mathrm{max}}$ to $\nu_{\mathrm{max}}$, the Doppler separation between different paths increases as $\nu_{\mathrm{max}}$ increases, and this reduces the inter-path interference. The slight increase beyond $\nu_{\max}=2000$~Hz is attributable to  
increased interference from the DD domain aliases (quasi-periodic replicas). For a given $\nu_{\mathrm{p}}$, increasing $\nu_{\mathrm{max}}$ will increase this interference as a consequence of increased overlap from the DD aliases. Also, TDL-A and TDL-C  achieve a lower BER than Veh-A by approximately one order of magnitude, again owing to their greater multipath diversity. The closeness between estimated CSI and perfect CSI plots demonstrates the robustness of the IOR estimation algorithm for large Dopplers. 
\section{Conclusions}
\label{sec:conclusion}
We considered the uplink of a multiuser Zak-OTFS system with heterogeneous users, where different users can be assigned different DD periods/frame sizes and user separation is achieved by different TF shifts to different users. A superimposed spread-pilot frame structure was considered for the estimation of IOR relation for each user. First,
we derived closed-form expressions for the effective DD domain channel between each user and the BS for sinc and Gaussian filters, using which  
we showed that the TF-shift based multiple access renders the IUI negligible, thereby decoupling the multiuser IOR estimation problem into independent single-user subproblems. Next, exploiting this decoupling, the IOR of each user is estimated independently at the BS using a DD dictionary-based approach. Simulation results for a multiuser 
heterogeneous system showed that the multiuser NMSE and BER performances closely match those of the corresponding single-user system, validating the user-decoupling approximation. Next, comparing the SE performance of embedded and superimposed pilot frames, we showed that superimposed frame achieves higher SE compared to embedded frame for sinc filter, whereas for Gaussian filter it is the reverse, i.e., embedded frame achieves higher SE compared to superimposed frame. We analyzed and explained the reason behind this interesting behavior. Finally, we investigated the robustness across Veh-A, TDL-A, and TDL-C channel models and maximum Dopplers, and demonstrated that the IOR estimation algorithm performs reliably under diverse and highly mobile channel conditions, with the gap between estimated CSI and perfect CSI performance curves remaining small throughout. Future work includes extension of the multiuser framework to multiantenna Zak-OTFS systems and optimal pilot-to-data and inter-user power allocation.

\appendices
\section{Derivation of (\ref{eq:heff_sinc}) and (\ref{eq:heff_gaussian})}
\label{app:heff_derivation}
Substituting the definition of twisted convolution,
$h_{\eff,u,v}(\tau,\nu)$ defined in (\ref{eq:ydd_filtered}) can be written as \cite{mu_paper}
\begin{eqnarray}
    h_{\eff,u,v}(\tau,\nu)
    & \hspace{-2mm} = & \hspace{-2mm}
    \sum_{i=1}^{P_v}
    h_{i,v}
    e^{j2\pi\nu_v(\tau-\tau_{i,v})}
    e^{j2\pi\nu_{i,v}(\tau+\tau_v-\tau_{i,v})} \notag\\
    & \hspace{-2mm} & \hspace{-2mm} 
    e^{-j2\pi\nu\tau_v}
    \zeta_{u,v,i}(\tau)
    \eta_{u,v,i}(\tau,\nu),
    \label{eqn:h_eff_mu_pap}
\end{eqnarray}
where
\begin{eqnarray}
\zeta_{u,v,i}(\tau)
& \hspace{-2mm} = & \hspace{-2mm}
\int
w^*_{B_u}(-\tau')
w_{B_v}(\tau-\tau'-\tau_{i,v}) \nonumber \\
& \hspace{-2mm} & \hspace{-2mm}
e^{j2\pi(\nu_u-\nu_v-\nu_{i,v})\tau'}
\,d\tau', 
\\
\eta_{u,v,i}(\tau,\nu)
& \hspace{-2mm} = & \hspace{-2mm}
\int
w^*_{T_u}(-\nu')
w_{T_v}(\nu-\nu'-\nu_{i,v}) \nonumber \\
& \hspace{-2mm} & \hspace{-2mm}
e^{j2\pi\nu'(\tau-\tau_u+\tau_v)}
\,d\nu'.
\end{eqnarray}
\subsection{Sinc filter}
\label{appx_A}
For the sinc filter, $\zeta_{u,v,i}(\tau)$ can be written as
\vspace{0mm}
\begin{align}
\zeta_{u,v,i}(\tau)
&= \sqrt{B_uB_v} \int \sinc(-B_u\tau') \sinc(B_v(\tau\hspace{-1mm}-\tau'\hspace{-1mm}-\tau_{i,v})) \nonumber \\
&\quad 
e^{j2\pi(\nu_u-\nu_v-\nu_{i,v})\tau'} \,d\tau' \nonumber \\
&= \sqrt{B_uB_v} \int \sinc(B_u\tau') \sinc(B_v(\tau'+\tau_{i,v}-\tau)) \nonumber \\
&\quad  
e^{j2\pi(\nu_u-\nu_v-\nu_{i,v})\tau'} \,d\tau' \nonumber \\
&\stackrel{(a)}{=} \frac{1}{\sqrt{B_uB_v}} \int \indic{-\frac{B_u}{2} \le \alpha-(\nu_u-\nu_v-\nu_{i,v}) < \frac{B_u}{2}} \nonumber \\
&\quad 
\indic{-\frac{B_v}{2} \le \alpha < \frac{B_v}{2}} e^{-j2\pi(\tau_{i,v}-\tau)\alpha} \,d\alpha \nonumber \\
&= \frac{\indic{\mathcal{B}_u\cap\mathcal{B}_v\neq\emptyset}}{\sqrt{B_uB_v}}\int_{b^{-}}^{b^{+}} e^{-j2\pi(\tau_{i,v}-\tau)\alpha} \,d\alpha \nonumber \\
&= \frac{\indic{\mathcal{B}_u\cap\mathcal{B}_v\neq\emptyset}}{\sqrt{B_uB_v}}  \frac{e^{-j2\pi(\tau_{i,v}-\tau)b^{-}} - e^{-j2\pi(\tau_{i,v}-\tau)b^{+}}}{j2\pi(\tau_{i,v}-\tau)} \nonumber \\
&= \frac{\indic{\mathcal{B}_u\cap\mathcal{B}_v\neq\emptyset}}{\sqrt{B_uB_v}}\frac{\sin\!\left(\pi(\tau_{i,v}-\tau)(b^{+}-b^{-})\right)}{\pi(\tau_{i,v}-\tau)} \nonumber \\
&\quad 
e^{-j\pi(\tau_{i,v}-\tau)(b^{-}+b^{+})} \nonumber \\
&= \frac{\indic{\mathcal{B}_u\cap\mathcal{B}_v\neq\emptyset}}{\sqrt{B_uB_v}}  (b^{+}\hspace{-1mm}-b^{-}) \sinc\!\left((\tau_{i,v}-\tau)(b^{+}\hspace{-1mm}-b^{-})\right) \nonumber \\
&\quad 
e^{-j\pi(\tau_{i,v}-\tau)(b^{-}+b^{+})},
\end{align}
where step $(a)$ is 
by using the Plancherel relation, and
\(
\mathcal{B}_u = \left[-\frac{B_u}{2}+(\nu_u-\nu_v-\nu_{i,v}), \, \frac{B_u}{2}+(\nu_u-\nu_v-\nu_{i,v})\right),
\mathcal{B}_v = \left[-\frac{B_v}{2}, \, \frac{B_v}{2}\right),
b^{-}=\inf(\mathcal{B}_u\cap\mathcal{B}_v),
b^{+}=\sup(\mathcal{B}_u\cap\mathcal{B}_v).
\)~Similarly, $\eta_{u,v,i}(\tau,\nu)$ can be obtained as
\begin{eqnarray}
\eta_{u,v,i}(\tau,\nu)
& \hspace{-3mm} = & \hspace{-3mm} \frac{\indic{\mathcal{T}_u\cap\mathcal{T}_v\neq\emptyset}}{\sqrt{T_uT_v}}  (t^{+}\hspace{-1mm}-t^{-}) \sinc\!\left((\nu_{i,v}\hspace{-0.5mm}-\nu)(t^{+}\hspace{-1mm}-t^{-})\right)
\nonumber \\
& \hspace{-2mm} & \hspace{-2mm}
e^{-j\pi(\nu_{i,v}-\nu)(t^{+}+t^{-})},
\end{eqnarray}
where
\(
\mathcal{T}_u = \left[-\frac{T_u}{2}+(\tau-\tau_u+\tau_v), \, \frac{T_u}{2}+(\tau-\tau_u+\tau_v)\right)\),
\(
\mathcal{T}_v = \left[-\frac{T_v}{2}, \, \frac{T_v}{2}\right),
t^{-}=\inf(\mathcal{T}_u\cap\mathcal{T}_v),
t^{+}=\sup(\mathcal{T}_u\cap\mathcal{T}_v)
\).
Substituting $\eta_{u,v,i}(\tau,\nu)$ and   $\zeta_{u,v,i}(\tau)$ in \eqref{eqn:h_eff_mu_pap}, $h_{\eff,u,v}(\tau,\nu)$ becomes \eqref{eq:heff_sinc}.

For the single-user case, $\tau_{i,v}=\tau_i$, $\nu_{i,v}=\nu_i$, $h_{i,v}=h_i$, $B_u=B_v=B$, $T_u=T_v=T$, $\tau_u=\tau_v=\nu_u=\nu_v=0$, the effective channel expression in (\ref{eq:heff_sinc}) specializes to
\begin{eqnarray}
h_{\eff}(\tau,\nu)
& \hspace{-2mm} = & \hspace{-2mm} \frac{1}{BT} \sum_{i=1}^{P} h_i e^{j\pi\nu_i(\tau-\tau_i)} e^{-j\pi(\nu_i-\nu)\tau} 
\nonumber \\
& \hspace{-12mm} & \hspace{-12mm} 
\sinc\!\left((\tau_i-\tau)(B-\abs{\nu_i})\right)\sinc\!\left((\nu_i-\nu)(T-\abs{\tau})\right) 
\nonumber \\
& \hspace{-12mm} & \hspace{-12mm}
\left(B-\abs{\nu_i}\right) \left(T-\abs{\tau}\right) \indic{\abs{\nu_i}<B} \indic{\abs{\tau}<T},
\end{eqnarray}
which when sampled at $\left(\tau=k\frac{\taup}{M},\nu=l\frac{\nup}{N}\right)$ becomes Eqn. (55) in \cite{closed_form}.

\subsection{Gaussian filter}
\label{appx_B}
For the Gaussian filter, $\zeta_{u,v,i}(\tau)$ can be written as
\begin{align}
&\zeta_{u,v,i}(\tau) \nonumber \\
&= \int w^*_{B_u}(-\tau') w_{B_v}(\tau-\tau'-\tau_{i,v}) e^{j2\pi(\nu_u-\nu_v-\nu_{i,v})\tau'} \,d\tau' \nonumber \\
&= \left(\frac{4\alpha_\tau^2B_u^2B_v^2}{\pi^2}\right)^{\frac14} \int e^{-\alpha_\tau B_u^2\tau'^2} e^{-\alpha_\tau B_v^2(\tau'-\tau+\tau_{i,v})^2} \nonumber \\
&\quad  
e^{j2\pi(\nu_u-\nu_v-\nu_{i,v})\tau'} \,d\tau'.
\end{align}
Defining
$a=\alpha_\tau B_u^2, b=\alpha_\tau B_v^2, c=\tau-\tau_{i,v}, f=\nu_u-\nu_v-\nu_{i,v}$, we write
\begin{align}
\zeta_{u,v,i}(\tau)
&= \left(\frac{4ab}{\pi^2}\right)^{\frac14} \int e^{-a\tau'^2} e^{-b(\tau'-c)^2} e^{j2\pi f\tau'} \,d\tau' \nonumber \\
&= \left(\frac{4ab}{\pi^2}\right)^{\frac14} e^{-bc^2} \int e^{-(a+b)\tau'^2} e^{(2bc+j2\pi f)\tau'} \,d\tau'.
\end{align}
Using
\begin{align}
\int_{-\infty}^{\infty} e^{-Ax^2+Bx}\,dx = \sqrt{\frac{\pi}{A}} e^{\frac{B^2}{4A}}, \qquad \Re(A)>0,    
\end{align}
we get
\begin{align}
\zeta_{u,v,i}(\tau)
&= \left(\frac{4ab}{\pi^2}\right)^{\frac14} \sqrt{\frac{\pi}{a+b}} e^{-\left(\frac{ab}{a+b}\right)c^2} e^{-\frac{\pi^2f^2}{a+b}} e^{j2\pi\left(\frac{bcf}{a+b}\right)}.
\end{align}
Substituting back the terms $a,b,~\text{and}~c$,
\begin{eqnarray}
\zeta_{u,v,i}(\tau)
& \hspace{-2mm} = & \hspace{-2mm} 
\left(\frac{4\alpha_\tau^2B_u^2B_v^2}{\pi^2}\right)^{\frac14} \!\!\!\sqrt{\frac{\pi}{\alpha_\tau(B_u^2+B_v^2)}} \nonumber \\
& \hspace{-2mm} & \hspace{-2mm}
e^{-\frac{\alpha_\tau B_u^2B_v^2}{B_u^2+B_v^2}(\tau-\tau_{i,v})^2} 
e^{-\frac{\pi^2(\nu_u-\nu_v-\nu_{i,v})^2}{\alpha_\tau(B_u^2+B_v^2)}} \nonumber \\
& \hspace{-2mm} & \hspace{-2mm}
e^{j2\pi\frac{B_v^2}{B_u^2+B_v^2}(\tau-\tau_{i,v})(\nu_u-\nu_v-\nu_{i,v})}.
\end{eqnarray}
Similarly,
\begin{eqnarray}
\eta_{u,v,i}(\tau,\nu)
& \hspace{-2mm} = & \hspace{-2mm}
\int w^*_{T_u}(-\nu') w_{T_v}(\nu-\nu'-\nu_{i,v}) \nonumber \\
& \hspace{-2mm} & \hspace{-2mm}
e^{j2\pi\nu'(\tau-\tau_u+\tau_v)}d\nu' \nonumber \\
& \hspace{-16mm} = & \hspace{-10mm}
\left(\frac{4\alpha_\nu^2T_u^2T_v^2}{\pi^2}\right)^{\frac14} \!\!\sqrt{\frac{\pi}{\alpha_\nu(T_u^2+T_v^2)}} e^{-\frac{\alpha_\nu T_u^2T_v^2}{T_u^2+T_v^2}(\nu-\nu_{i,v})^2} \nonumber \\
& \hspace{-16mm} & \hspace{-10mm}  
e^{-\frac{\pi^2(\tau-\tau_u+\tau_v)^2}{\alpha_\nu(T_u^2+T_v^2)}} e^{j2\pi\frac{T_v^2}{T_u^2+T_v^2}(\nu-\nu_{i,v})(\tau-\tau_u+\tau_v)}.
\end{eqnarray}
Substituting $\eta_{u,v,i}(\tau,\nu)$ and   $\zeta_{u,v,i}(\tau)$ in \eqref{eqn:h_eff_mu_pap}, $h_{\eff,u,v}(\tau,\nu)$ becomes \eqref{eq:heff_gaussian}.

For the single-user case, the effective channel expression in (\ref{eq:heff_gaussian}) specializes to
\begin{eqnarray}
h_{\eff}(\tau,\nu)
& \hspace{-2mm} = & \hspace{-2mm} 
\sum_{i=1}^{P} h_i e^{j2\pi\nu_i(\tau-\tau_i)} e^{-\frac{\alpha_\tau B^2}{2}(\tau-\tau_i)^2} e^{-\frac{\pi^2\nu_i^2}{2\alpha_\tau B^2}} \nonumber \\
& \hspace{-2mm} & \hspace{-2mm} 
e^{-\frac{\alpha_\nu T^2}{2}(\nu-\nu_i)^2} e^{-\frac{\pi^2\tau^2}{2\alpha_\nu T^2}} e^{-j\pi(\tau-\tau_i)\nu_i} e^{j\pi(\nu-\nu_i)\tau} \nonumber \\
& \hspace{-2mm} = & \hspace{-2mm}
\sum_{i=1}^{P} h_i e^{-\frac{\alpha_\tau B^2}{2}(\tau-\tau_i)^2} e^{-\frac{\pi^2\nu_i^2}{2\alpha_\tau B^2}}e^{-\frac{\alpha_\nu T^2}{2}(\nu-\nu_i)^2} \nonumber \\
& \hspace{-2mm} & \hspace{-2mm}  
e^{-\frac{\pi^2\tau^2}{2\alpha_\nu T^2}} e^{j\pi\nu_i(\tau-\tau_i)} e^{j\pi\tau(\nu-\nu_i)} \nonumber \\
& \hspace{-2mm} = & \hspace{-2mm}
\sum_{i=1}^{P} h_i e^{-\left(\frac{\alpha_\tau B^2}{2}(\tau-\tau_i)^2 +\frac{\alpha_\nu T^2}{2}(\nu-\nu_i)^2\right)} \nonumber \\
& \hspace{-2mm} & \hspace{-2mm}   
e^{-\left(\frac{\pi^2\nu_i^2}{2\alpha_\tau B^2} +\frac{\pi^2\tau^2}{2\alpha_\nu T^2}\right)} e^{j\pi\big(\tau\nu - \tau_i\nu_i\big)},
\end{eqnarray}
which when sampled at $\left(\tau=k\frac{\taup}{M},\nu=l\frac{\nup}{N}\right)$ becomes Eqn. (62) in \cite{closed_form}.

\section{Derivation of (\ref{eqn:sinc_noise_cov1}) and (\ref{eqn:gauss_noise_cov1}) }
\label{app:noise_stat}
Substituting $w_{\mathrm{tx}}(\tau,\nu) = w_{Bu}(\tau)\,w_{Tu}(\nu)\,e^{j2\pi(\nu_u\tau-\tau_u\nu)}$, $w_{\mathrm{rx}}(\tau,\nu) = w_{\mathrm{tx}}^{*}(-\tau,-\nu)\,e^{j2\pi\tau\nu}$, the receive filter is $w_{\mathrm{rx}}(\tau,\nu) = w_{Bu}^{*}(-\tau)\,w_{Tu}^{*}(-\nu)\,e^{j2\pi(\nu_u\tau-\tau_u\nu)}\,e^{j2\pi\tau\nu}$, and by definition of Zak-transform \cite{zakotfs_pred}, $n_{\mathrm{dd},u}(\tau,\nu)$ can be written as
\begin{align}
n_{\mathrm{dd},u}(\tau,\nu) = \sqrt{\tau_{\mathrm{p},u}}\sum_{q\in\mathbb{Z}} n(\tau+q\tau_{\mathrm{p},u})\,e^{-j2\pi q\frac{\nu}{\nu_{\mathrm{p},u}}}.    
\end{align}
The filtered DD domain noise is
\begin{align}
&n_{\mathrm{dd},u}^{w_{\mathrm{rx}}}(\tau,\nu) \nonumber \\
&= \iint w_{\mathrm{rx}}(\tau',\nu')\,n_{\mathrm{dd},u}(\tau-\tau',\nu-\nu')e^{j2\pi\nu'(\tau-\tau')}\,d\tau'd\nu'\notag \\
&= \iint w_{Bu}^{*}(-\tau')\,w_{Tu}^{*}(-\nu')e^{j2\pi(\nu_u\tau'-\tau_u\nu')}e^{j2\pi\tau'\nu'} \notag\\
&\quad
e^{j2\pi\nu'(\tau-\tau')}  
\hspace{-1mm}\sqrt{\tau_{\mathrm{p},u}}\sum_{q\in\mathbb{Z}} n(\tau\hspace{-1mm}-\tau'\hspace{-1mm}+q\tau_{\mathrm{p},u})e^{-j2\pi q\frac{(\nu-\nu')}{\nu_{\mathrm{p},u}}}\,d\tau'd\nu'\notag \\
&= \sqrt{\tau_{\mathrm{p},u}}\sum_{q\in\mathbb{Z}} e^{-j2\pi q\frac{\nu}{\nu_{\mathrm{p},u}}}
   \iint_{\tau',\nu'} w_{Bu}^{*}(-\tau')\,w_{Tu}^{*}(-\nu')e^{j2\pi\nu'\tau} \notag\\
&\quad 
n(\tau-\tau'+q\tau_{\mathrm{p},u})e^{j2\pi(\nu_u\tau'-\tau_u\nu')}e^{j2\pi q\frac{\nu'}{\nu_{\mathrm{p},u}}}\,d\tau'\,d\nu'\notag \\
&= \sqrt{\tau_{\mathrm{p},u}}\sum_{q\in\mathbb{Z}} e^{-j2\pi q\frac{\nu}{\nu_{\mathrm{p},u}}}\hspace{-1mm}\underbrace{\int_{\nu'} \hspace{-1mm} w_{Tu}^{*}(-\nu') 
   e^{j2\pi\nu'\left(\frac{q}{\nu_{\mathrm{p},u}}+\tau-\tau_u\right)}\,d\nu'}_{\displaystyle I_1(\tau)}
   \notag \\
&\quad 
\int_{\tau'} w_{Bu}^{*}(-\tau')\,n(\tau-\tau'+q\tau_{\mathrm{p},u})e^{j2\pi\nu_u\tau'}\,d\tau'.
\label{eqn:noise_cov}
\end{align}

\subsection{Sinc filter}
\label{appx_C}
For the sinc filter, substituting $w_{Bu}(\tau)=\sqrt{B_u}\,\sinc(B_u\tau)$ and $w_{Tu}(\nu)=\sqrt{T_u}\,\sinc(T_u\nu)$, $I_1(\tau)$ in (\ref{eqn:noise_cov}) can be written as
\begin{align}
I_1(\tau)
&= \sqrt{T_u}\int_{\nu'} \sinc(T_u\nu')\,
   e^{j2\pi\nu'\left(\frac{q}{\nu_{\mathrm{p},u}}+\tau-\tau_u\right)}\,d\nu' \nonumber \\
&= \frac{1}{\sqrt{T_u}}\;\indic{ -\frac{T_u}{2}\le \frac{q}{\nu_{\mathrm{p},u}}+\tau-\tau_u < \frac{T_u}{2}}.
\label{eqn:I1s}
\end{align}
Substituting  
(\ref{eqn:I1s}) 
in \eqref{eqn:noise_cov}, we get
\begin{eqnarray}
n_{\mathrm{dd},u}^{w_{\mathrm{rx}}}(\tau,\nu) 
& \hspace{-0mm} & \hspace{-0mm} \nonumber \\
& \hspace{-22mm} = & \hspace{-12mm} \sqrt{\tfrac{B_u\tau_{\mathrm{p},u}}{T_u}}\sum_{q\in\mathbb{Z}}
   e^{-j2\pi q\frac{\nu}{\nu_{\mathrm{p},u}}}
   \indic{ -\frac{T_u}{2}\le \frac{q}{\nu_{\mathrm{p},u}}+\tau-\tau_u < \frac{T_u}{2} }
    \nonumber \\
& \hspace{-22mm} & \hspace{-12mm} 
\int_{\tau'} \sinc(B_u\tau')\,n(\tau-\tau'+q\tau_{\mathrm{p},u})\,e^{j2\pi\nu_u\tau'}\,d\tau' \nonumber \\
& \hspace{-22mm} = & \hspace{-12mm} \sqrt{\tfrac{B_u\,\tau_{\mathrm{p},u}}{T_u}}
   \sum_{q=\left\lceil -\frac{N_u}{2}-\frac{\tau}{\tau_{\mathrm{p},u}}+\frac{\tau_u}{\tau_{\mathrm{p},u}}\right\rceil}
        ^{\left\lfloor \frac{N_u}{2}-\frac{\tau}{\tau_{\mathrm{p},u}}+\frac{\tau_u}{\tau_{\mathrm{p},u}}\right\rfloor}
   e^{-j2\pi q\frac{\nu}{\nu_{\mathrm{p},u}}} \nonumber \\
& \hspace{-22mm} & \hspace{-12mm}
   \int_{\tau'} \sinc(B_u\tau')\,n(\tau-\tau'+q\tau_{\mathrm{p},u})\,e^{j2\pi\nu_u\tau'}\,d\tau'.
   \label{eqn:samps}
\end{eqnarray}
Sampling (\ref{eqn:samps}) at the DD grid points
$\tau=\frac{k\tau_{\mathrm{p},u}}{M_u}$, $\nu=\frac{l\nu_{\mathrm{p},u}}{N_u}$, we get
\begin{eqnarray}
\hspace{-6mm}
n_{\mathrm{dd},u}^{w_{\mathrm{rx}}}[k,l]
& \hspace{-2mm} = & \hspace{-2mm} \sqrt{\tfrac{B_u\tau_{\mathrm{p},u}}{T_u}}
   \sum_{q=\left\lceil -\frac{N_u}{2}-\frac{k}{M_u}+\frac{\tau_u}{\tau_{\mathrm{p},u}}\right\rceil}
        ^{\left\lfloor \frac{N_u}{2}-\frac{k}{M_u}+\frac{\tau_u}{\tau_{\mathrm{p},u}}\right\rfloor}
   e^{-j2\pi q\frac{l}{N_u}} \notag\\
& \hspace{-10mm} & \hspace{-10mm}\int_{\tau'} \sinc(B_u\tau')\,n(k\tfrac{\tau_{\p,u}}{M_u}-\tau'+q\tau_{\mathrm{p},u})\,e^`{j2\pi\nu_u\tau'}\,d\tau'.
\label{eqn:noise_sampled_sinc}
\end{eqnarray}
Using \eqref{eqn:noise_sampled_sinc}, the covariance between $n_{\mathrm{dd},u}^{w_{\mathrm{rx}}}[k_1,l_1]$ and $n_{\mathrm{dd},u}^{w_{\mathrm{rx}}}[k_2,l_2]$ can be written as
\begin{eqnarray}
\mathbb{E}\left[ n_{\mathrm{dd},u}^{w_{\mathrm{rx}}}[k_1,l_1]\; n_{\mathrm{dd},u}^{w_{\mathrm{rx}}*}[k_2,l_2]\,\right] & & \notag \\
& \hspace{-88mm} = & \hspace{-45mm} \tfrac{B_u\tau_{\mathrm{p},u}}{T_u}\!\!\!\!\!\!\!\!\!\!\!\!
   \sum_{q_1=\left\lceil -\frac{N_u}{2}-\frac{k_1}{M_u}+\frac{\tau_u}{\tau_{\mathrm{p},u}}\right\rceil}
        ^{\left\lfloor \frac{N_u}{2}-\frac{k_1}{M_u}+\frac{\tau_u}{\tau_{\mathrm{p},u}}\right\rfloor}
           \sum_{q_2=\left\lceil -\frac{N_u}{2}-\frac{k_2}{M_u}+\frac{\tau_u}{\tau_{\mathrm{p},u}}\right\rceil}
        ^{\left\lfloor \frac{N_u}{2}-\frac{k_2}{M_u}+\frac{\tau_u}{\tau_{\mathrm{p},u}}\right\rfloor}
   \!\!\!\!\!\!\!\!\!\!\!\!e^{-j2\pi\frac{q_1 l_1}{N_u}}e^{j2\pi\frac{q_2 l_2}{N_u}} \notag\\
& \hspace{-45mm} & \hspace{-45mm}  
\mathbb{E}\Bigg[\iint_{\tau_1',\tau_2'} \sinc(B_u\tau_1')\,\sinc(B_u\tau_2')e^{j2\pi\nu_u(\tau_1'-\tau_2')} \notag \\
& \hspace{-45mm} & \hspace{-45mm} n\left(\tfrac{k_1\tau_{\mathrm{p},u}}{M_u}-\tau_1'+q_1\tau_{\mathrm{p},u}\right)
   n^{*}\!\left(\tfrac{k_2\tau_{\mathrm{p},u}}{M_u}-\tau_2'+q_2\tau_{\mathrm{p},u}\right) \nonumber \\
   & \hspace{-45mm} & \hspace{-45mm} 
   d\tau_1'd\tau_2'\Bigg].
   \label{eqn:sinc_exp_term_1}
\end{eqnarray}
The expectation term in \eqref{eqn:sinc_exp_term_1} can be solved as
\begin{eqnarray}
N_0\iint_{\tau_1',\tau_2'} \sinc(B_u\tau_1')\,\sinc(B_u\tau_2')e^{j2\pi\nu_u(\tau_1'-\tau_2')} & & \nonumber \\
& \hspace{-75mm} & \hspace{-75mm}
   \delta\!\left(\tfrac{k_1\tau_{\mathrm{p},u}}{M_u}-\tau_1'+q_1\tau_{\mathrm{p},u}
   -\Big(\tfrac{k_2\tau_{\mathrm{p},u}}{M_u}-\tau_2'+q_2\tau_{\mathrm{p},u}\Big)\right)
   \,d\tau_1'd\tau_2' \nonumber \\
& \hspace{-140mm} = & \hspace{-71mm} N_0\iint_{\tau_1',\tau_2'} \sinc(B_u\tau_1')\,\sinc(B_u\tau_2')\,
      e^{j2\pi\nu_u(\tau_1'-\tau_2')} \nonumber \\
& \hspace{-140mm} & \hspace{-71mm}
\delta\!\left(\tfrac{(k_1-k_2)\tau_{\mathrm{p},u}}{M_u}-(\tau_1'-\tau_2')+(q_1-q_2)\tau_{\mathrm{p},u}\right)\,d\tau_1'\,d\tau_2' \nonumber \\
& \hspace{-140mm} = & \hspace{-71mm} N_0e^{j2\pi\nu_u\left(\frac{(k_1-k_2)\tau_{\mathrm{p},u}}{M_u}+(q_1-q_2)\tau_{\mathrm{p},u}\right)} 
\int_{\tau_1'} \sinc(B_u\tau_1')
\nonumber \\
& \hspace{-140mm} & \hspace{-71mm}
\sinc\left(B_u\left(\tau_1'-(k_1-k_2)\tfrac{\tau_{\mathrm{p},u}}{M_u}-(q_1-\!q_2)\tau_{\mathrm{p},u}\right)\right)d\tau_1' \nonumber \\
& \hspace{-140mm} = & \hspace{-71mm} 
\frac{N_0}{B_u} e^{j2\pi\nu_u\left(\frac{(k_1-k_2)\tau_{\mathrm{p},u}}{M_u}+(q_1-q_2)\tau_{\mathrm{p},u}\right)} \nonumber \\
& \hspace{-140mm} & \hspace{-71mm} \sinc\left(B_u\Big((k_2-k_1)\tfrac{\tau_{\mathrm{p},u}}{M_u}+(q_2-q_1)\tau_{\mathrm{p},u}\Big)\right).
\end{eqnarray}
Substituting the above expression for the expectation in \eqref{eqn:sinc_exp_term_1}, the covariance becomes
\begin{eqnarray}
\mathbb{E}\left[\, n_{\mathrm{dd},u}^{w_{\mathrm{rx}}}[k_1,l_1]\; n_{\mathrm{dd},u}^{w_{\mathrm{rx}}*}[k_2,l_2]\,\right] & & \nonumber \\
& \hspace{-75mm} = & \hspace{-38mm} N_0 \frac{\tau_{\mathrm{p},u}}{T_u} 
   \sum_{q_1=\left\lceil -\frac{N_u}{2}-\frac{k_1}{M_u}+\frac{\tau_u}{\tau_{\mathrm{p},u}}\right\rceil}
        ^{\left\lfloor \frac{N_u}{2}-\frac{k_1}{M_u}+\frac{\tau_u}{\tau_{\mathrm{p},u}}\right\rfloor}
   \sum_{q_2=\left\lceil -\frac{N_u}{2}-\frac{k_2}{M_u}+\frac{\tau_u}{\tau_{\mathrm{p},u}}\right\rceil}
        ^{\left\lfloor \frac{N_u}{2}-\frac{k_2}{M_u}+\frac{\tau_u}{\tau_{\mathrm{p},u}}\right\rfloor}
   \nonumber \\
& \hspace{-75mm} & \hspace{-38mm}   
e^{\,j\frac{2\pi}{N_u}(q_2 l_2-q_1 l_1)}e^{j2\pi\nu_u\left(\frac{(k_1-k_2)\tau_{\mathrm{p},u}}{M_u}+(q_1-q_2)\tau_{\mathrm{p},u}\right)} \nonumber \\
& \hspace{-75mm} & \hspace{-38mm} 
\sinc\!\left(B_u\Big((k_2-k_1)\tfrac{\tau_{\mathrm{p},u}}{M_u}+(q_2-q_1)\tau_{\mathrm{p},u}\Big)\right).
\end{eqnarray}

\subsection{Gaussian filter}
\label{appx_D}
For the Gaussian filter, substituting~$w_{B_u}(\tau)=\left(\frac{2\alpha_\tau B_u^2}{\pi}\right)^{1/4}e^{-\alpha_\tau B_u^2\tau^2}$ and
$w_{T_u}(\nu)=\left(\frac{2\alpha_\nu T_u^2}{\pi}\right)^{1/4}
e^{-\alpha_\nu T_u^2\nu^2}$, $I_1(\tau)$  in (\ref{eqn:noise_cov}) becomes
\begin{eqnarray}
I_1(\tau)
& \hspace{-2mm} = & \hspace{-2mm} \left(\frac{2\alpha_\nu T_u^2}{\pi}\right)^{\!1/4}
\int_{\nu'} e^{-\alpha_\nu T_u^2\nu'^2}
e^{j2\pi\nu'\!\left(\frac{q}{\nu_{\mathrm{p},u}}+\tau-\tau_u\right)}
d\nu' \nonumber \\
& \hspace{-2mm} = & \hspace{-2mm} \left(\frac{2\alpha_\nu T_u^2}{\pi}\right)^{\!1/4}
\sqrt{\frac{\pi}{\alpha_\nu T_u^2}}\,
e^{-\frac{\pi^2}{\alpha_\nu T_u^2}
\!\left(\frac{q}{\nu_{\mathrm{p},u}}+\tau-\tau_u\right)^{\!2}} \nonumber \\
& \hspace{-2mm} = & \hspace{-2mm} \left(\frac{2\pi}{\alpha_\nu T_u^2}\right)^{\!1/4}
e^{-\frac{\pi^2}{\alpha_\nu T_u^2}
\!\left(\frac{q}{\nu_{\mathrm{p},u}}+\tau-\tau_u\right)^{\!2}}.
\label{eqn:I1g}
\end{eqnarray}
Substituting 
(\ref{eqn:I1g}) in \eqref{eqn:noise_cov}, we get
\begin{eqnarray}
n_{\dd,u}^{w_{\mathrm{rx}}}(\tau,\nu)
& \hspace{-2mm} = & \hspace{-2mm} 
\left(\frac{4\alpha_\tau B_u^2}{\alpha_\nu T_u^2}
\right)^{\!1/4}\!\sqrt{\tau_{\mathrm{p},u}}
\sum_{q\in\mathbb{Z}}
e^{-j2\pi q\frac{\nu}{\nu_{\mathrm{p},u}}} \nonumber \\
& \hspace{-2mm} & \hspace{-2mm}
e^{-\frac{\pi^2}{\alpha_\nu T_u^2}
\!\left(\frac{q}{\nu_{\mathrm{p},u}}+\tau-\tau_u\right)^{\!2}}
\int_{\tau'}
e^{-\alpha_\tau B_u^2\tau'^2} \nonumber \\
& \hspace{-2mm} & \hspace{-2mm}
n(\tau-\tau'+q\tau_{\mathrm{p},u})\,
e^{j2\pi\nu_u\tau'}\,d\tau'.
\label{eqn:sampg}
\end{eqnarray}
Sampling (\ref{eqn:sampg}) at the DD grid points
$\tau=\frac{k\tau_{\mathrm{p},u}}{M_u}$, $\nu=\frac{l\nu_{\mathrm{p},u}}{N_u}$, we get
\begin{eqnarray}
n_{\dd,u}^{w_{\mathrm{rx}}}[k,l]
& \hspace{-2mm} = & \hspace{-2mm} \left(\frac{4\alpha_\tau B_u^2}{\alpha_\nu T_u^2}
\right)^{\!1/4}\!\sqrt{\tau_{\mathrm{p},u}}
\sum_{q\in\mathbb{Z}}
e^{-j2\pi q\frac{l}{N_u}} \notag \\
& \hspace{-2mm} & \hspace{-2mm}
e^{-\frac{\pi^2}{\alpha_\nu T_u^2}
\!\left(\frac{q}{\nu_{\mathrm{p},u}}
+\frac{k\tau_{\mathrm{p},u}}{M_u}-\tau_u\right)^{\!2}}
\int_{\tau'}
e^{-\alpha_\tau B_u^2\tau'^2} \notag \\
& \hspace{-2mm} & \hspace{-2mm}
n\!\left(\tfrac{k\tau_{\mathrm{p},u}}{M_u}
-\tau'+q\tau_{\mathrm{p},u}\right)
e^{j2\pi\nu_u\tau'}\,d\tau'.
\label{eqn:gauss_sampled_noise}
\end{eqnarray}
Using \eqref{eqn:gauss_sampled_noise}, the covariance between 
$n_{\mathrm{dd},u}^{w_{\text{rx}}}[k_1,l_1]$ and $n_{\mathrm{dd},u}^{w_{\text{rx}}}[k_2,l_2]$ can be written as
\begin{eqnarray}
\mathbb{E}\left[
n_{\dd,u}^{w_{\mathrm{rx}}}[k_1,l_1]\,
n_{\dd,u}^{w_{\mathrm{rx}}*}[k_2,l_2]
\right] & & \nonumber \\
& \hspace{-80mm} = & \hspace{-40mm}
N_0\sqrt{\frac{4\alpha_\tau B_u^2}{\alpha_\nu T_u^2}}
\tau_{\mathrm{p},u}
\sum_{q_1\in\mathbb{Z}}
\sum_{q_2\in\mathbb{Z}}
e^{-j\frac{2\pi}{N_u}(q_1l_1-q_2l_2)} 
\nonumber \\
& \hspace{-80mm} & \hspace{-40mm} 
e^{-\frac{\pi^2}{\alpha_\nu T_u^2}
\left(
\frac{q_1}{\nu_{\mathrm{p},u}}
+\frac{k_1\tau_{\mathrm{p},u}}{M_u}
-\tau_u
\right)^2} 
e^{-\frac{\pi^2}{\alpha_\nu T_u^2}
\left(
\frac{q_2}{\nu_{\mathrm{p},u}}
+\frac{k_2\tau_{\mathrm{p},u}}{M_u}
-\tau_u
\right)^2}
\nonumber \\
& \hspace{-80mm} & \hspace{-40mm} 
\int_{\tau'}
e^{-\alpha_\tau B_u^2\tau'^2}
e^{-\alpha_\tau B_u^2(\tau'+\Delta\tau)^2}
e^{j2\pi\nu_u\Delta\tau} d\tau',
\end{eqnarray}
where
$\Delta\tau
=
\frac{(k_1-k_2)\tau_{\mathrm{p},u}}{M_u}
+
(q_1-q_2)\tau_{\mathrm{p},u}.$
Using the property
\begin{equation}
\int_{-\infty}^{\infty}
e^{-\alpha_\tau B_u^2\tau'^2}
e^{-\alpha_\tau B_u^2(\tau'+\Delta\tau)^2}
d\tau'
 = 
\sqrt{\frac{\pi}{2\alpha_\tau B_u^2}}
e^{-\frac{\alpha_\tau B_u^2}{2}(\Delta\tau)^2},
\end{equation}
we obtain
\begin{eqnarray}
\mathbb{E}\left[
n_{\dd,u}^{w_{\mathrm{rx}}}[k_1,l_1]\,
n_{\dd,u}^{w_{\mathrm{rx}}*}[k_2,l_2]
\right] & & \nonumber \\
& \hspace{-80mm} = & \hspace{-41mm} N_0
\sqrt{\frac{2\pi}{\alpha_\nu T_u^2}}\,
\tau_{\mathrm{p},u}
\sum_{q_1\in\mathbb{Z}}
\sum_{q_2\in\mathbb{Z}}
e^{-j\frac{2\pi}{N_u}(q_1l_1-q_2l_2)}
\nonumber \\
& \hspace{-80mm} & \hspace{-41mm} 
e^{-\frac{\pi^2}{\alpha_\nu T_u^2}
\left(
\frac{q_1}{\nu_{\mathrm{p},u}}
+\frac{k_1\tau_{\mathrm{p},u}}{M_u}
-\tau_u
\right)^2}
e^{-\frac{\pi^2}{\alpha_\nu T_u^2}
\left(
\frac{q_2}{\nu_{\mathrm{p},u}}
+\frac{k_2\tau_{\mathrm{p},u}}{M_u}
-\tau_u
\right)^2}
\nonumber \\
& \hspace{-80mm} & \hspace{-41mm}
e^{j2\pi\nu_u\Delta\tau}
e^{-\frac{\alpha_\tau B_u^2}{2}
(\Delta\tau)^2}.
\end{eqnarray}
For the single-user case, the covariance expressions for sinc and Gaussian specialize to Eqns. (61) and (63) in \cite{closed_form}, respectively.


\begin{thebibliography}{99}
\bibitem{next_gen_ITU}
``Framework and overall objectives of the future development of IMT for 2030 and beyond,'' ITU-R M.2160-0, Nov. 2023. https://www.itu.int/rec/R-REC-M.2160/en.

\bibitem{otfs_WCNC}
R. Hadani et al., ``Orthogonal time frequency space modulation,'' \textit{Proc. IEEE WCNC'2017}, pp. 1-6, Mar. 2017.

\bibitem{SE}
I. A. Khan et al., ``Does 6G need a new waveform:
Comparing Zak-OTFS with CP-OFDM,'' arXiv:2601.15602, Jan. 2026.

\bibitem{zakotfs_bits}
S. K. Mohammed, R. Hadani, A. Chockalingam, and
R. Calderbank, ``OTFS -- A mathematical foundation for communication and radar sensing in the delay-Doppler domain,'' \textit{IEEE BITS Inf. Theory Mag.}, vol. 2, no. 2, pp. 36-55, 1 Nov. 2022.

\bibitem{zakotfs_pred}
S. K. Mohammed, R. Hadani, A. Chockalingam, and
R. Calderbank, ``OTFS -- Predictability in the
delay-Doppler domain and its value in communication and radar sensing,'' \textit{IEEE BITS Inf. Theory Mag.}, vol. 3, no. 2, pp. 7-31, Jun. 2023.

\bibitem{mc_otfs_1}
K. R. Murali and A. Chockalingam, ``On OTFS modulation for high-Doppler fading channels,'' {\em Proc. ITA'2018}, pp. 1-10, Feb. 2018.

\bibitem{mc_otfs_2} 
P. Raviteja, K. T. Phan, Y. Hong, and E. Viterbo, ``Interference cancellation and iterative detection for orthogonal time frequency space modulation,'' {\em IEEE Trans. Wireless Commun.,} vol. 17, no. 10, pp. 6501-6515, Oct. 2018.

\bibitem{mu_otfs_1}
V. Khammammetti and S. K. Mohammed, ``OTFS-based multiple-access in high Doppler and delay spread wireless channels,'' \textit{IEEE Wireless Commun. Lett.}, vol. 8, no. 2, pp. 528-531, Apr. 2019.

\bibitem{mu_otfs_2}
V. Khammammetti and S. K. Mohammed, ``Spectral efficiency of OTFS based orthogonal multiple access with rectangular pulses,'' \textit{IEEE Trans. Veh. Tech.}, vol. 71, no. 12, pp. 12989-13006, Dec. 2022.

\bibitem{mu_otfs_3}
S. Rakib and R. Hadani, ``Multiple access in wireless telecommunications system for high-mobility applications,” \textit{U.S. Patent} 9 722 741 B1, Aug. 2017.

\bibitem{mu_otfs_4}
R. M. Augustine and A. Chockalingam, ``Interleaved time-frequency multiple access using OTFS modulation,`` \textit{Proc. IEEE VTC’2019-Fall}, pp. 1–5, Sep. 2019.

\bibitem{mu_otfs_6}
O. K. Rasheed, G. D. Surabhi, and A. Chockalingam,
``Sparse delay-Doppler channel estimation in rapidly
time-varying channels for multiuser OTFS on the uplink,'' \textit{Proc. IEEE VTC'2020-Spring}, May 2020.

\bibitem{mu_otfs_7}
M. Li, S. Zhang, P. Fan, and O. A. Dobre, ``Multiple access for massive MIMO-OTFS networks over angle-delay-Doppler domain,'' \textit{Proc. IEEE GLOBECOM’2020}, pp. 1–6, Dec. 2020.

\bibitem{mu_otfs_8}
M. Li, S. Zhang, F. Gao, P. Fan and, O. A. Dobre, ``A new path division multiple access for the
massive MIMO-OTFS networks,'' \textit{IEEE J. Sel. Areas Commun.}, vol. 39, no. 4, pp. 903–918, Apr.
2021.

\bibitem{Zak_existance}
A. J. E. M. Janssen, ``Bargmann transform, Zak transform, and coherent states,'' \textit{J. Math. Phys.} 23, 720–731 (1982).

\bibitem{zak_transform}
A. J. E. M. Janssen, ``The Zak transform: a signal transform for sampled time-continuous signals,'' {\it Philips J. Res.}, 43, pp. 23-69, 1988.

\bibitem{closed_form}
A. Das, F. Jesbin, and A. Chockalingam,
``Closed-form expressions for I/O relation in Zak-OTFS with different delay-Doppler filters,'' \textit{IEEE Trans. Veh. Tech..}, vol. 74, no. 9, pp. 14250--14266, Sep. 2025.

\bibitem{gsfilter}
A. Das, F. Jesbin, and A. Chockalingam, ``A Gaussian-sinc pulse shaping filter for Zak-OTFS,''
\textit{IEEE Trans. Veh. Tech.}, vol. 75, no. 3, pp. 4307-4322, Mar. 2026.

\bibitem{isac}
M. Ubadah, S. K. Mohammed, R. Hadani, S. Kons,
A. Chockalingam, and R. Calderbank, ``Zak-OTFS to
integrate sensing the I/O relation and data
communication,'' arXiv:2404.04182, Jun. 2025.

\bibitem{zadoff_spread}
S. R. Mattu, I. A. Khan, V. Khammammetti, B. Dabak, S. K. Mohammed, K. Narayanan, and R. Calderbank, ``Delay-Doppler signal processing with Zadoff-Chu sequences,'' arXiv:2412.04295, Dec. 2024.

\bibitem{Fathima_VTC}
F. Jesbin, J. Bodempudi, and A. Chockalingam, ``Diversity and BER performance of Zak-OTFS modulation,'' \textit{Proc. IEEE VTC2025-Spring}, pp. 1-6, Jun. 2025.

\bibitem{goldsmith}
A. Goldsmith, \textit{Wireless Communications}, Cambridge Univ. Press, 2005.

\bibitem{mu_paper}
I. A. Khan, S. K. Mohammed, R. Hadani, A. Chockalingam, and R. Calderbank, ``Zak-OTFS based multiuser uplink in doubly-spread channels,'' arXiv:2507.15621, Jul. 2025.

\bibitem{VehA}
ITU-R M.1225, ``Guidelines for evaluation of radio
transmission technologies for IMT-2000,''
Int. Telecom. Union Radio Commun., 1997.

\bibitem{tdl}
3GPP TR 38.901, ``Study on channel model for frequencies from 0.5 to 100 GHz,'' version 16.1.0, Release 16.

\bibitem{lsmr}
H. Qu, G. Liu, L. Zhang, S. Wen, and M. A. Imran,
``Low-complexity symbol detection and interference
cancellation for OTFS system,'' \textit{IEEE Trans.
Commun.}, vol. 69, no. 3, pp. 1524-1537, Mar. 2021.

\end{thebibliography}
\end{document}